\input harvmac
\input epsf

%&plain
%
\let\includefigures=\iftrue
%
% the following is to use blackboard bold fonts --
%\let\useblackboard=\iftrue
%
% activate this if you don't have them.
%\let\useblackboard=\iffalse
%
% You might also need to remove this line.
\newfam\black
\input rotate
\input epsf
\noblackbox
%
%draftmode
%
\includefigures
\message{If you do not have epsf.tex (to include figures),}
\message{change the option at the top of the tex file.}
\def\figin{\epsfcheck\figin}\def\figins{\epsfcheck\figins}
\def\epsfcheck{\ifx\epsfbox\UnDeFiNeD
\message{(NO epsf.tex, FIGURES WILL BE IGNORED)}
\gdef\figin##1{\vskip2in}\gdef\figins##1{\hskip.5in}% blank space instead
\else\message{(FIGURES WILL BE INCLUDED)}%
\gdef\figin##1{##1}\gdef\figins##1{##1}\fi}
\def\DefWarn#1{}
\def\N{{\cal N}}
\def\figinsert{\goodbreak\midinsert}
\def\ifig#1#2#3{\DefWarn#1\xdef#1{fig.~\the\figno}
\writedef{#1\leftbracket fig.\noexpand~\the\figno}%
\figinsert\figin{\centerline{#3}}\medskip\centerline{\vbox{\baselineskip12pt
\advance\hsize by -1truein\noindent\footnotefont{\bf
Fig.~\the\figno:} #2}}
\bigskip\endinsert\global\advance\figno by1}
%%%
\else
\def\ifig#1#2#3{\xdef#1{fig.~\the\figno}
\writedef{#1\leftbracket fig.\noexpand~\the\figno}%
%\figinsert\figin{\centerline{#3}}\medskip\centerline{\vbox{\baselineskip12pt
%\advance\hsize by -1truein\noindent\footnotefont{\bf Fig.~\the\figno:} #2}}
%\bigskip\endinsert
\global\advance\figno by1} \fi

\def\tilde{\widetilde}

\def\yboxit#1#2{\vbox{\hrule height #1 \hbox{\vrule width #1
\vbox{#2}\vrule width #1 }\hrule height #1 }}
\def\fillbox#1{\hbox to #1{\vbox to #1{\vfil}\hfil}}
\def\ybox{{\lower 1.3pt \yboxit{0.4pt}{\fillbox{8pt}}\hskip-0.2pt}}

\def\rightarrowbox#1#2{
  \setbox1=\hbox{\kern#1{${ #2}$}\kern#1}
  \,\vbox{\offinterlineskip\hbox to\wd1{\hfil\copy1\hfil}
    \kern 3pt\hbox to\wd1{\rightarrowfill}}}

\def\half{{1\over 2}}
\def\Tr{{{\rm Tr~ }}}

\def\vev#1{\langle{#1}\rangle}

\def\tilde{\widetilde}

\def\II{\relax{I\kern-.10em I}}

\def\IZ{\relax\ifmmode\mathchoice
{\hbox{\cmss Z\kern-.4em Z}}{\hbox{\cmss Z\kern-.4em Z}}
{\lower.9pt\hbox{\cmsss Z\kern-.4em Z}} {\lower1.2pt\hbox{\cmsss
Z\kern-.4em Z}}\else{\cmss Z\kern-.4em Z}\fi}
\def\IB{\relax{\rm I\kern-.18em B}}
\def\IC{{\relax\hbox{$\inbar\kern-.3em{\rm C}$}}}
\def\ID{\relax{\rm I\kern-.18em D}}
\def\IE{\relax{\rm I\kern-.18em E}}
\def\IF{\relax{\rm I\kern-.18em F}}
\def\IG{\relax\hbox{$\inbar\kern-.3em{\rm G}$}}
\def\IGa{\relax\hbox{${\rm I}\kern-.18em\Gamma$}}
\def\IH{\relax{\rm I\kern-.18em H}}
\def\II{\relax{\rm I\kern-.18em I}}
\def\IK{\relax{\rm I\kern-.18em K}}
\def\IN{\relax{\rm I\kern-.18em N}}
\def\IP{\relax{\rm I\kern-.18em P}}
%\def\IX{\relax{\rm X\kern-.01em X}}
%this doesn't work

%
\def\inbar{\,\vrule height1.5ex width.4pt depth0pt}

\font\cmss=cmss10 \font\cmsss=cmss10 at 7pt
\def\IR{\relax{\rm I\kern-.18em R}}

\def\lp10{l_P^{10}}
\def\lp11{l_P^{11}}
\def\R11{R_{11}}

\def\gb#1{ {\langle #1 ] } }

\newbox\tmpbox\setbox\tmpbox\hbox{\abstractfont
}
 \Title{\vbox{\baselineskip12pt\hbox to\wd\tmpbox{\hss
 hep-th/0410179} }}
 {\vbox{\centerline{Computing One-Loop Amplitudes From}
 \bigskip
 \centerline{ The Holomorphic Anomaly Of Unitarity Cuts  }
 }}
\smallskip
\centerline{Ruth Britto, Freddy Cachazo, and Bo Feng}
\smallskip
\bigskip
\centerline{\it School of Natural Sciences, Institute for Advanced
Study, Princeton NJ 08540 USA}
\bigskip
\vskip 1cm \noindent

\input amssym.tex

We propose a systematic way to carry out the method introduced in
hep-th/0410077 for computing certain unitarity cuts of one-loop
${\cal N}=4$ amplitudes of gluons. We observe that the class of
cuts for which the method works involves all next-to-MHV n-gluon
one-loop amplitudes of any helicity configurations. As an
application of our systematic procedure, we obtain the complete
seven-gluon one-loop leading-color amplitude
$A_{7;1}(1^-,2^-,3^-,4^+,5^+,6^+,7^+)$.

\Date{October 2004}

\lref\BernZX{ Z.~Bern, L.~J.~Dixon, D.~C.~Dunbar and
D.~A.~Kosower, ``One Loop N Point Gauge Theory Amplitudes,
Unitarity And Collinear Limits,'' Nucl.\ Phys.\ B {\bf 425}, 217
(1994), hep-ph/9403226.
%%CITATION = HEP-PH 9403226;%%
}

%\BernCG
\lref\BernCG{ Z.~Bern, L.~J.~Dixon, D.~C.~Dunbar and
D.~A.~Kosower, ``Fusing Gauge Theory Tree Amplitudes into Loop
Amplitudes,'' Nucl.\ Phys.\ B {\bf 435}, 59 (1995),
hep-ph/9409265.
%%CITATION = HEP-PH 9409265;%%
}

%\WittenNN
\lref\WittenNN{ E.~Witten, ``Perturbative Gauge Theory as a String
Theory in Twistor Space,'' hep-th/0312171.
%%CITATION = HEP-TH 0312171;%%
}

%\CachazoKJ
\lref\CachazoKJ{ F.~Cachazo, P.~Svr\v cek and E.~Witten, ``MHV
Vertices and Tree Amplitudes in Gauge Theory,'' hep-th/0403047.
%%CITATION = HEP-TH 0403047;%%
}

\lref\berkwitten{N. Berkovits and E. Witten,  ``Conformal
Supergravity In Twistor-String Theory,'' hep-th/0406051.}

\lref\penrose{R. Penrose, ``Twistor Algebra,'' J. Math. Phys. {\bf
8} (1967) 345.}

\lref\berends{F. A. Berends, W. T. Giele and H. Kuijf, ``On
Relations Between Multi-Gluon And Multi-Graviton Scattering,"
Phys. Lett {\bf B211} (1988) 91.}

\lref\berendsgluon{F. A. Berends, W. T. Giele and H. Kuijf,
``Exact and Approximate Expressions for Multigluon Scattering,"
Nucl. Phys. {\bf B333} (1990) 120.}

\lref\bernplusa{Z. Bern, L. Dixon and D. A. Kosower, ``New QCD
Results From String Theory,'' in {\it Strings '93}, ed. M. B.
Halpern et. al. (World-Scientific, 1995), hep-th/9311026.}

\lref\bernplusb{Z. Bern, G. Chalmers, L. J. Dixon and D. A.
Kosower, ``One Loop $N$ Gluon Amplitudes with Maximal Helicity
Violation via Collinear Limits," Phys. Rev. Lett. {\bf 72} (1994)
2134.}

\lref\bernfive{Z. Bern, L. J. Dixon and D. A. Kosower, ``One Loop
Corrections to Five Gluon Amplitudes," Phys. Rev. Lett {\bf 70}
(1993) 2677.}

\lref\bernfourqcd{Z.Bern and  D. A. Kosower, "The Computation of
Loop Amplitudes in Gauge Theories," Nucl. Phys.  {\bf B379,}
(1992) 451.}

\lref\cremmerlag{E. Cremmer and B. Julia, ``The $N=8$ Supergravity
Theory. I. The Lagrangian," Phys. Lett.  {\bf B80} (1980) 48.}

\lref\cremmerso{E. Cremmer and B. Julia, ``The $SO(8)$
Supergravity," Nucl. Phys.  {\bf B159} (1979) 141.}

\lref\dewitt{B. DeWitt, "Quantum Theory of Gravity, III:
Applications of Covariant Theory," Phys. Rev. {\bf 162} (1967)
1239.}

\lref\dunbarn{D. C. Dunbar and P. S. Norridge, "Calculation of
Graviton Scattering Amplitudes Using String Based Methods," Nucl.
Phys. B {\bf 433,} 181 (1995), hep-th/9408014.}

\lref\ellissexton{R. K. Ellis and J. C. Sexton, "QCD Radiative
corrections to parton-parton scattering," Nucl. Phys.  {\bf B269}
(1986) 445.}

\lref\gravityloops{Z. Bern, L. Dixon, M. Perelstein, and J. S.
Rozowsky, ``Multi-Leg One-Loop Gravity Amplitudes from Gauge
Theory,"  hep-th/9811140.}

\lref\kunsztqcd{Z. Kunszt, A. Singer and Z. Tr\'{o}cs\'{a}nyi,
``One-loop Helicity Amplitudes For All $2\rightarrow2$ Processes
in QCD and ${\cal N}=1$ Supersymmetric Yang-Mills Theory,'' Nucl.
Phys.  {\bf B411} (1994) 397, hep-th/9305239.}

\lref\mahlona{G. Mahlon, ``One Loop Multi-photon Helicity
Amplitudes,'' Phys. Rev.  {\bf D49} (1994) 2197, hep-th/9311213.}

\lref\mahlonb{G. Mahlon, ``Multi-gluon Helicity Amplitudes
Involving a Quark Loop,''  Phys. Rev.  {\bf D49} (1994) 4438,
hep-th/9312276.}

\lref\klt{H. Kawai, D. C. Lewellen and S.-H. H. Tye, ``A Relation
Between Tree Amplitudes of Closed and Open Strings," Nucl. Phys.
{B269} (1986) 1.}

\lref\pppmgr{Z. Bern, D. C. Dunbar and T. Shimada, ``String Based
Methods In Perturbative Gravity," Phys. Lett.  {\bf B312} (1993)
277, hep-th/9307001.}

%\GiombiIX
\lref\GiombiIX{ S.~Giombi, R.~Ricci, D.~Robles-Llana and
D.~Trancanelli, ``A Note on Twistor Gravity Amplitudes,''
hep-th/0405086.
%%CITATION = HEP-TH 0405086;%%
}

%\WuFB
\lref\WuFB{ J.~B.~Wu and C.~J.~Zhu, ``MHV Vertices and Scattering
Amplitudes in Gauge Theory,'' hep-th/0406085.
%%CITATION = HEP-TH 0406085;%%
}

\lref\Feynman{R.P. Feynman, Acta Phys. Pol. 24 (1963) 697, and in
{\it Magic Without Magic}, ed. J. R. Klauder (Freeman, New York,
1972), p. 355.}

\lref\Peskin{M.~E. Peskin and D.~V. Schroeder, {\it An Introduction
to Quantum Field Theory} (Addison-Wesley Pub. Co., 1995).}

\lref\parke{S. Parke and T. Taylor, ``An Amplitude For $N$ Gluon
Scattering,'' Phys. Rev. Lett. {\bf 56} (1986) 2459; F. A. Berends
and W. T. Giele, ``Recursive Calculations For Processes With $N$
Gluons,'' Nucl. Phys. {\bf B306} (1988) 759. }

\lref\BrandhuberYW{ A.~Brandhuber, B.~Spence and G.~Travaglini,
``One-Loop Gauge Theory Amplitudes In N = 4 Super Yang-Mills From
MHV Vertices,'' hep-th/0407214.
%%CITATION = HEP-TH 0407214;%%
}

\lref\CachazoZB{ F.~Cachazo, P.~Svr\v cek and E.~Witten, ``Twistor
space structure of one-loop amplitudes in gauge theory,''
hep-th/0406177.
%%CITATION = HEP-TH 0406177;%%
}

\lref\passarino{ L.~M. Brown and R.~P. Feynman, ``Radiative Corrections To Compton Scattering,'' Phys. Rev. 85:231
(1952); G.~Passarino and M.~Veltman, ``One Loop Corrections For E+ E- Annihilation Into Mu+ Mu- In The Weinberg
Model,'' Nucl. Phys. B160:151 (1979);
G.~'t Hooft and M.~Veltman, ``Scalar One Loop Integrals,'' Nucl. Phys. B153:365 (1979); R.~G.~
Stuart, ``Algebraic Reduction Of One Loop Feynman Diagrams To Scalar Integrals,'' Comp. Phys. Comm. 48:367 (1988); R.~G.~Stuart and A.~Gongora, ``Algebraic Reduction Of One Loop Feynman Diagrams To Scalar Integrals. 2,'' Comp. Phys. Comm. 56:337 (1990).}

\lref\neerven{ W. van Neerven and J. A. M. Vermaseren, ``Large Loop Integrals,'' Phys. Lett.
137B:241 (1984)}

\lref\melrose{ D.~B.~Melrose, ``Reduction Of Feynman Diagrams,'' Il Nuovo Cimento 40A:181 (1965); G.~J.~van Oldenborgh and J.~A.~M.~Vermaseren, ``New Algorithms For One Loop Integrals,'' Z. Phys. C46:425 (1990);
G.J. van Oldenborgh,  PhD Thesis, University of Amsterdam (1990);
A. Aeppli, PhD thesis, University of Zurich (1992).}

\lref\bernTasi{Z.~Bern, hep-ph/9304249, in {\it Proceedings of
Theoretical Advanced Study Institute in High Energy Physics (TASI
92)}, eds. J. Harvey and J. Polchinski (World Scientific, 1993). }

%\BernTZ
\lref\morgan{ Z.~Bern and A.~G.~Morgan, ``Supersymmetry relations
between contributions to one loop gauge boson amplitudes,'' Phys.\
Rev.\ D {\bf 49}, 6155 (1994), hep-ph/9312218.
%%CITATION = HEP-PH 9312218;%%
}

\lref\RoiSpV{R.~Roiban, M.~Spradlin and A.~Volovich, ``A Googly
Amplitude From The B-Model In Twistor Space,'' JHEP {\bf 0404},
012 (2004) hep-th/0402016; R.~Roiban and A.~Volovich, ``All Googly
Amplitudes From The $B$-Model In Twistor Space,'' hep-th/0402121;
R.~Roiban, M.~Spradlin and A.~Volovich, ``On The Tree-Level
S-Matrix Of Yang-Mills Theory,'' Phys.\ Rev.\ D {\bf 70}, 026009
(2004) hep-th/0403190.}

%\CachazoBY
\lref\CachazoBY{ F.~Cachazo, P.~Svr\v cek and E.~Witten, ``Gauge
Theory Amplitudes In Twistor Space And Holomorphic Anomaly,''
hep-th/0409245.
%%CITATION = HEP-TH 0409245;%%
}

%\DixonWI
\lref\DixonWI{ L.~J.~Dixon, ``Calculating Scattering Amplitudes
Efficiently,'' hep-ph/9601359.
%%CITATION = HEP-PH 9601359;%%
}

%\BernMQ
\lref\BernMQ{ Z.~Bern, L.~J.~Dixon and D.~A.~Kosower, ``One Loop
Corrections To Five Gluon Amplitudes,'' Phys.\ Rev.\ Lett.\  {\bf
70}, 2677 (1993), hep-ph/9302280.
%%CITATION = HEP-PH 9302280;%%
}

\lref\berends{F.~A.~Berends, R.~Kleiss, P.~De Causmaecker, R.~Gastmans and T.~T.~Wu, ``Single Bremsstrahlung Processes In Gauge Theories,'' Phys. Lett. {\bf B103} (1981) 124; P.~De
Causmaeker, R.~Gastmans, W.~Troost and T.~T.~Wu, ``Multiple Bremsstrahlung In Gauge Theories At High-Energies. 1. General
Formalism For Quantum Electrodynamics,'' Nucl. Phys. {\bf
B206} (1982) 53; R.~Kleiss and W.~J.~Stirling, ``Spinor Techniques For Calculating P Anti-P $\to$ W+- / Z0 + Jets,'' Nucl. Phys. {\bf
B262} (1985) 235; R.~Gastmans and T.~T. Wu, {\it The Ubiquitous
Photon: Heliclity Method For QED And QCD} Clarendon Press, 1990.}

\lref\xu{Z. Xu, D.-H. Zhang and L. Chang, ``Helicity Amplitudes For Multiple
Bremsstrahlung In Massless Nonabelian Theories,''
 Nucl. Phys. {\bf B291}
(1987) 392.}

\lref\gunion{J.~F. Gunion and Z. Kunszt, ``Improved Analytic Techniques For Tree Graph Calculations And The G G Q
Anti-Q Lepton Anti-Lepton Subprocess,''
Phys. Lett. {\bf 161B}
(1985) 333.}

%\GeorgiouBY
\lref\GeorgiouBY{ G.~Georgiou, E.~W.~N.~Glover and V.~V.~Khoze,
``Non-MHV Tree Amplitudes In Gauge Theory,'' JHEP {\bf 0407}, 048
(2004), hep-th/0407027.
%%CITATION = HEP-TH 0407027;%%
}

%\WuJX
\lref\WuJX{ J.~B.~Wu and C.~J.~Zhu, ``MHV Vertices And Fermionic
Scattering Amplitudes In Gauge Theory With Quarks And Gluinos,''
hep-th/0406146.
%%CITATION = HEP-TH 0406146;%%
}

%\WuFB
\lref\WuFB{ J.~B.~Wu and C.~J.~Zhu, ``MHV Vertices And Scattering
Amplitudes In Gauge Theory,'' JHEP {\bf 0407}, 032 (2004),
hep-th/0406085.
%%CITATION = HEP-TH 0406085;%%
}

%\GeorgiouWU
\lref\GeorgiouWU{ G.~Georgiou and V.~V.~Khoze, ``Tree Amplitudes
In Gauge Theory As Scalar MHV Diagrams,'' JHEP {\bf 0405}, 070
(2004), hep-th/0404072.
%%CITATION = HEP-TH 0404072;%%
}

\lref\Nair{V. Nair, ``A Current Algebra For Some Gauge Theory
Amplitudes," Phys. Lett. {\bf B78} (1978) 464. }

%\BernAD
\lref\BernAD{ Z.~Bern, ``String Based Perturbative Methods For
Gauge Theories,'' hep-ph/9304249.
%%CITATION = HEP-PH 9304249;%%
}

%\BernKR
\lref\BernKR{ Z.~Bern, L.~J.~Dixon and D.~A.~Kosower,
``Dimensionally Regulated Pentagon Integrals,'' Nucl.\ Phys.\ B
{\bf 412}, 751 (1994), hep-ph/9306240.
%%CITATION = HEP-PH 9306240;%%
}

%\CachazoDR
\lref\CachazoDR{ F.~Cachazo, ``Holomorphic Anomaly Of Unitarity
Cuts And One-Loop Gauge Theory Amplitudes,'' hep-th/0410077.
%%CITATION = HEP-TH 0410077;%%
}

\lref\giel{W. T. Giele and E. W. N. Glover, ``Higher order corrections to jet cross-sections in e+ e- annihilation,'' Phys. Rev. {\bf D46}
(1992) 1980; W. T. Giele, E. W. N. Glover and D. A. Kosower, ``Higher order corrections to jet cross-sections in hadron colliders,'' Nucl.
Phys. {\bf B403} (1993) 633. }

\lref\kuni{Z. Kunszt and D. Soper, ``Calculation of jet cross-sections in hadron collisions at order alpha-s**3,''Phys. Rev. {\bf D46} (1992)
192; Z. Kunszt, A. Signer and Z. Tr\' ocs\' anyi, ``Singular terms of helicity amplitudes at one loop in QCD and the soft limit
of the cross-sections of multiparton processes,'' Nucl. Phys. {\bf
B420} (1994) 550. }

\lref\seventree{F.~A. Berends, W.~T. Giele and H. Kuijf, ``Exact And Approximate Expressions For Multi - Gluon Scattering,'' Nucl. Phys.
{\bf B333} (1990) 120.}

\lref\mangpxu{M. Mangano, S.~J. Parke and Z. Xu, ``Duality And Multi - Gluon Scattering,'' Nucl. Phys. {\bf B298}
(1988) 653.}

\lref\mangparke{M. Mangano and S.~J. Parke, ``Multiparton Amplitudes In Gauge Theories,'' Phys. Rep. {\bf 200}
(1991) 301.}

\lref\grisaru{M. T. Grisaru, H. N. Pendleton and P. van Nieuwenhuizen, ``Supergravity And The S Matrix,'' Phys. Rev.  {\bf D15} (1977) 996; M. T. Grisaru and H. N. Pendleton, ``Some Properties Of Scattering Amplitudes In Supersymmetric Theories,'' Nucl. Phys. {\bf B124} (1977) 81.}

\lref\Bena{I. Bena, Z. Bern, D. A. Kosower and R. Roiban, ``Loops in Twistor Space,'' hep-th/0410054.}

%\BernKY
\lref\BernKY{
Z.~Bern, V.~Del Duca, L.~J.~Dixon and D.~A.~Kosower,
``All Non-Maximally-Helicity-Violating One-Loop Seven-Gluon Amplitudes In N =
4 Super-Yang-Mills Theory,''
arXiv:hep-th/0410224.
%%CITATION = HEP-TH 0410224;%%
}

%%%%%%%%%%%%%%%%%%%%%%%%%%%%%%%%%%%%%
\newsec{Introduction}

One-loop amplitudes of gluons in supersymmetric gauge theories
possess many remarkable properties. One of them is that they are
four-dimensional cut constructible \refs{\BernZX, \BernCG}. This
means that the amplitudes are completely determined by their
unitarity cuts.

Recently, a new method for computing certain unitarity cuts of
one-loop amplitudes in $\N=4$ gauge theories was proposed in
\CachazoDR. The method uses the fact that unitarity cuts can be
computed in two ways.

One is by a cut integral, where two tree-level amplitudes are
connected by cut propagators. The other is by computing the
imaginary part of the amplitude in a certain kinematical regime
chosen in order to isolate the given cut.

In general, the amplitudes of interest are not known. However,
they can be written as linear combinations of scalar box functions
with unknown rational coefficients in the kinematical
variables\foot{This is strictly true in the spinor-helicity
formalism of \refs{\berends,\xu,\gunion}.}
\refs{\passarino,\neerven,\melrose}. These functions are
completely known in terms of logarithms and dilogarithms \BernKR.

The key observation made in \CachazoDR\ is that if a given first-order differential operator acts on the cut integral to produce a
rational function, then the operator must annihilate the
coefficients that multiply the scalar box functions in the
amplitude. This ensures that the result of applying the operator
to the imaginary part of the amplitude is also a rational
function.

The problem of finding the unknown coefficients in the amplitude
is thus related to that of comparing two rational functions.

The rational function obtained from the action of the operator on
the imaginary part of the amplitude naturally comes out as a sum
over ``simple fractions". On the other hand, the rational function
that comes from the action of the operator on the cut integral
comes out in a compact form.

The aim of this paper is to provide a systematic method for
carrying out the reduction of the latter into the form of the
former. Once this is done, the unknown coefficients in the
amplitude can simply be read off by directly comparing the two
expressions.

In \CachazoDR, a simple prescription was given for finding suitable
operators for cuts where at least one of the tree-level amplitudes
in the cut integral representation is maximally helicity
violating (MHV). The idea is that when amplitudes are
transformed to twistor space, they are localized on simple
algebraic sets \WittenNN. In particular, MHV tree-level amplitudes
are localized on lines. In \WittenNN, differential operators for
testing the localization of gluons on lines (collinear operators)
were introduced. By using the holomorphic anomaly of unitarity
cuts found in \CachazoBY\ by combining the results of \CachazoZB\ and \BrandhuberYW, one can prove that these operators can
only produce rational functions when acting on the cut
integrals \CachazoDR.

We also find that all unitarity cuts of next-to-MHV n-gluon
one-loop amplitudes of any helicity configuration satisfy the
requirements to be computable by our method. This extends the
class of amplitudes given in \CachazoDR\ from
$A_{n;1}(1^-,2^-,3^-,4^+,\ldots, n^+)$ to amplitudes with three
negative helicity gluons in arbitrary positions.

One-loop amplitudes of gluons that are known explicitly are very
rare. The largest set is known for $\N=4$ amplitudes, where all
n-gluon MHV amplitudes are known \BernZX. In addition to this
series of amplitudes, only the six-gluon next-to-MHV one-loop
amplitude with any helicity configuration is known \BernCG.

In this paper, we illustrate    our general method by calculating the
seven-gluon next-to-MHV amplitude
$A_{7;1}(1^-,2^-,3^-,4^+,5^+,6^+,7^+)$. This calculation involves
the computation of the coefficients of thirty-five scalar box
functions. This is the first amplitude where the three-mass scalar
box function participates.

This paper is organized as follows. In section 2, we explain the
systematic reduction procedure that produces the coefficients of
the scalar box functions in the amplitude. In section 3, we apply
our general method to the calculation of the seven-gluon amplitude
$A_{7;1}(1^-,2^-,3^-,4^+,5^+,6^+,7^+)$. In section 4, we write
down the explicit form of the coefficient of the thirty-five
scalar box functions that participate in the seven-gluon
amplitude. In Appendix A, we give the explicit form of the
scalar box functions and discuss their infrared singular behavior.
In Appendix B, we prove that we our method gives complete information about all next-to-MHV amplitudes with any helicity configuration.

Throughout the paper, we use the following notation and conventions.  The external gluon labeled by $i$ carries momentum $p_i$.  
\eqn\ournot{
\eqalign{
s_{ij} &\equiv 2 p_i \cdot p_j = \vev{i~j} [i~j], \cr t_i^{[r]} &\equiv
(p_i+p_{i+1}+\cdots+p_{i+r-1})^2, \cr \gb{i|j_r+j_{r+1}+\cdots
+j_s|k} &\equiv
\vev{i~j_r}[j_r~k]+\vev{i~j_{r+1}}[j_{r+1}~k]+\cdots+\vev{i~j_s}[i~j_s].
}}

{\bf Note added in third version:}

The reader will be interested to know that the seven-gluon amplitude with the helicity configuration $(---++++)$ has now also been computed in \BernKY, along with all other helicity configurations, using the direct unitarity method.  
(Please be warned that the first version of our paper contained a typo in the coefficient $d_{3,4}$ and a corresponding typo in $d_{2,3},$ which was obtained by a permutation of labels.)  
It is interesting to note that, according to \BernKY, the reduction techniques of the direct unitarity method give ``quite large'' formulas for the coefficients.
One advantage of our method is that we derive the coefficients analytically in a simple form.  
The authors of \BernKY\ were able to produce similarly simple formulas by postulating ans\"atze that were checked numerically at random kinematic points.

%%%%%%%%%%%%%%%%%%%%%%%%%%%%%%%%%

\newsec{General Reduction Techniques}

One-loop amplitudes of gluons in supersymmetric gauge theories are
four dimensional-cut constructible. This means that knowing the
discontinuities of the amplitude is enough to fix the amplitude
completely \BernZX. Having QCD computations in mind, one should
consider one-loop amplitudes in $\N=4$ super Yang-Mills as well as
one-loop amplitudes with an $\N=1$ chiral super multiplet running
in the loop.

Even though we concentrate on $\N=4$ amplitudes, it should be kept
in mind that everything is valid, with some minor modifications,
for $\N=1$ amplitudes.

The problem at hand is the computation of the leading-color
n-gluon one-loop $\N=4$ amplitudes. This is the part of the full
amplitude proportional to $N \Tr (T^{a_1}\ldots T^{a_n})$.

These amplitudes can be written as linear combinations of scalar
box functions, which are listed explicitly in Appendix A.
 (For $\N=1$ one also has to include scalar triangle
and bubble functions.)
\eqn\gene{ A_{n;1}^{\rm 1-loop} = \sum_{i=1}^n \left( b_i
F^{1m}_{n:i}+ \sum_r c_{r,i} F^{2m~e}_{n:r;i}+ \sum_r d_{r,i}
F^{2m~h}_{n:r;i}+ \sum_{r,r'} g_{ r,r',i} F^{3m}_{n:r:r';i}
\right).}
 This
means that computing the amplitude is equivalent to computing the
coefficients. Note that we have not included four-mass scalar box
functions. The reason is that for the classes of amplitudes
considered in this paper these cannot appear, as proven in
\CachazoDR.

A new technique to compute these coefficients was proposed in
\CachazoDR. The basic idea is to compute the unitarity cuts of
\gene\ using the holomorphic anomaly found in \CachazoBY. Here we
present a systematic procedure to carry out the proposal of
\CachazoDR\ that is directly applicable to all cuts of next-to-MHV
one-loop amplitudes.

\ifig\convi{Representation of the cut integral. Left and right
tree-level amplitudes are on-shell. Internal lines represent the
legs coming from the cut propagators.}
{\epsfxsize=0.50\hsize\epsfbox{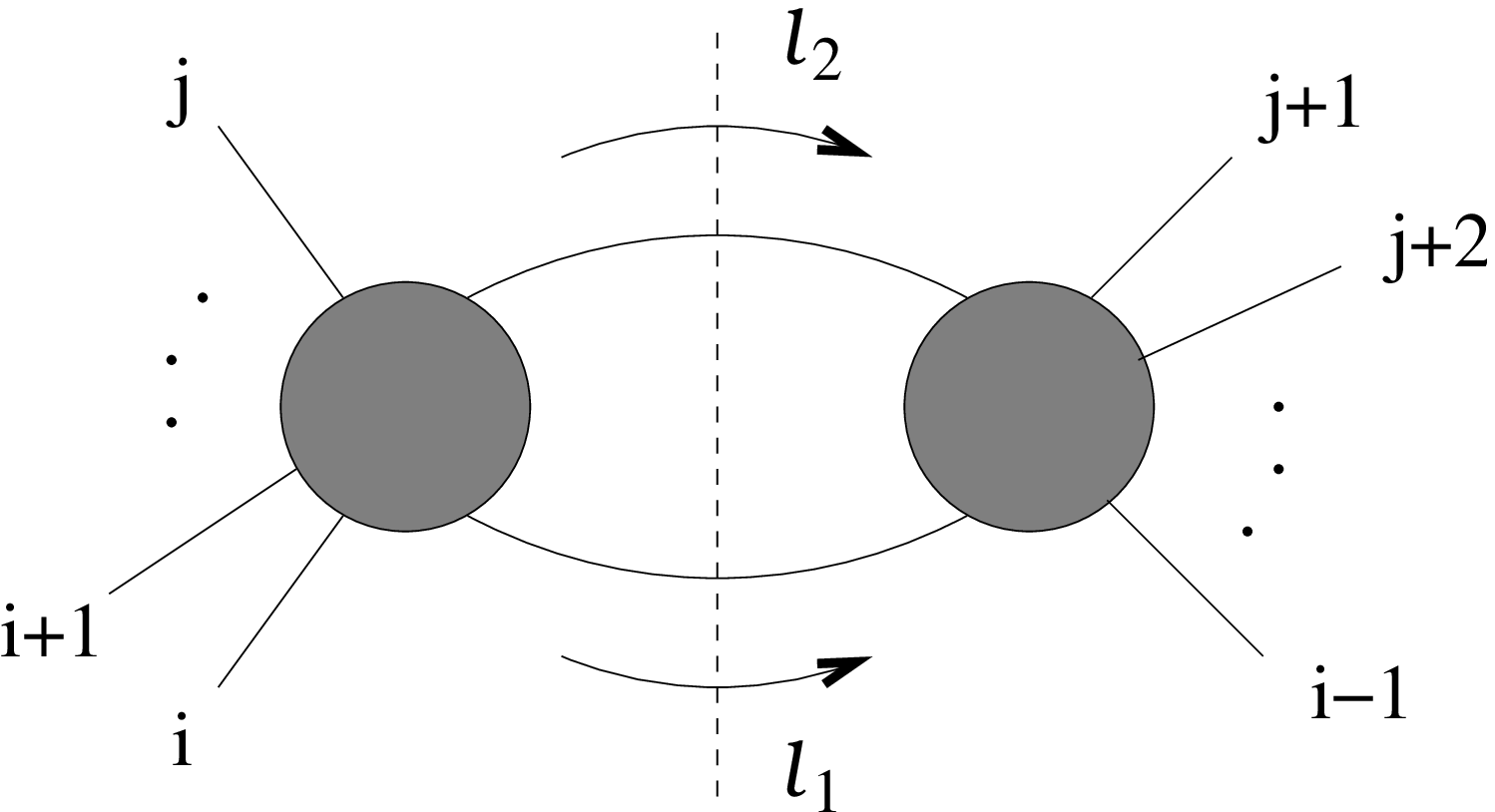}}

Consider the unitarity cut in the $(i,i+1,\ldots ,
j-1,j)$-channel. This is given by the cut integral
\eqn\cutIn{ \eqalign{ & C_{i,i+1,\ldots ,j-1,j} = \cr &   \int
d\mu A^{\rm tree}((-\ell_1),i,i+1,\ldots ,j-1,j,(-\ell_2))A^{\rm
tree}(\ell_2,j+1,j+2,\ldots ,i-2,i-1,\ell_1),}}
where $d\mu$ is the Lorentz invariant phase space measure of two
light-like vectors $(\ell_1, \ell_2)$ constrained by momentum
conservation. We find it useful to define $\ell_1$ and $\ell_2$ as
in \convi. We follow the conventions of \CachazoDR.

This cut can also be computed by the taking the imaginary part of
the full amplitude in the kinematical regime where $t_i^{[j-i+1]}
= (p_i+p_{i+1}+\ldots +p_j)^2$ is positive and all other
invariants are negative \BernZX.

It is now clear that computing $C_{i,i+1,\ldots ,j-1,j}$ provides
information about the amplitude via
\eqn\word{C_{i,i+1,\ldots ,j-1,j} = {\rm Im}|_{t_{i}^{[j-i+1]}>0}
A_{n;1}.}

The class of cuts considered in \CachazoDR\ are those for which
one of the tree-level amplitudes in \cutIn\ is an MHV amplitude.
All next-to-MHV amplitudes have this property.  If all three negative-helicity gluons appear on the same side of the cut, then the amplitude on the other side of the cut either vanishes or is MHV.  If one side of the cut has exactly one negative-helicity gluon, there are three cases to consider for the helicities of the cut propagators on this side.  If they are both positive, this tree amplitude vanishes.  If exactly one is positive, then it is MHV.  If both are negative, then their helicities are positive viewed from the other side of the cut, so that side is the MHV amplitude. 

Let the left tree-level amplitude in \cutIn\ be the MHV amplitude
\parke\
\eqn\treeMHV{A^{\rm tree MHV}_{km}((-\ell_1),i,(i+1),\ldots
,j,(-\ell_2)) = {\vev{k~m}^4\over \vev{\ell_1~ i}\vev{i~i+1}\cdots
\vev{j-1~j}\vev{j~\ell_2}\vev{\ell_2~\ell_1}}.}
Using this in \cutIn\ we have
\eqn\didi{\eqalign{ & C_{i,i+1,\ldots ,j-1,j} = \cr & \int d\mu
{\vev{k~m}^4\over \vev{i~i+1}\ldots
\vev{j-1~j}\vev{\ell_2~\ell_1}} {1\over
\vev{\ell_1~i}\vev{j~\ell_2}}A^{\rm tree}(\ell_2,j+1,j+2,\ldots
,i-1,\ell_1) .}  }

The basic idea is to find a differential operator of first order
that produces a rational function when acting on the cut \didi.
Let ${\cal O}$ be such an operator. Then ${\cal O}C_{i,i+1,\ldots
,j-1,j}$ is a rational function. A simple prescription for finding
such operators and for computing the rational function explicitly
was given in \CachazoDR. We postpone this for the moment; we
do not need the explicit form of the operator in what follows.

Consider now the action of ${\cal O}$ on \word, i.e.,
\eqn\weri{{\cal O}C_{i,i+1,\ldots ,j-1,j} = {\cal O}~{\rm
Im}|_{t_{i}^{[j-i+1]}>0} A_{n;1}. }
Since the operator ${\cal O}$ is of first order, it produces two
terms for each term in the amplitude \gene:  one term when it acts
on the scalar box function and one more when it acts on the
coefficient. It turns out that the imaginary part of each scalar
box function is the logarithm of a rational function $R$ of the
kinematical invariants.\foot{This is not true for the four-mass
scalar box function, but as proven in \CachazoDR\ these do not
contribute to the cuts we consider.} Therefore, when ${\cal O}$
acts on the logarithms it produces rational functions. However,
when it acts on the coefficients, the logarithms survive. In
\CachazoDR\ it was proven that the only way this can be consistent
with the fact that ${\cal O}C_{i,i+1,\ldots ,j-1,j}$ is a rational
function is that ${\cal O}$ annihilates the coefficients.

This means that we can write ${\cal O}C_{i,i+1,\ldots ,j-1,j}$
schematically as follows:
\eqn\scheme{{\cal O}C_{i,i+1,\ldots ,j-1,j} = \sum_{k} a_k {{\cal
O}( R_k) \over R_k},}
where $a_k$ stands for a general coefficient in \gene, and the sum runs over the terms produced by all box functions
that develop an imaginary part in the kinematical regime of
interest for this cut.

Now we can clearly describe the mathematical problem involved in
the calculations of the coefficients $a_k$.

{}From the action of the operator on the cut integral we find a
rational function
\eqn\ratt{{\cal O}C_{i,i+1,\ldots ,j-1,j} = {P\over Q \prod_k G_k},
}
where $P$, $Q$ and $G_k$ are polynomials. Generically $P$ is not
annihilated by ${\cal O}$. On the other hand, we have defined $Q$
such that ${\cal O}Q = 0$. All other factors in the denominator
that are not annihilated by ${\cal O}$ become one of the $G_k$.

The problem is to find a way of writing \ratt\ in the form
\scheme\ in order to read off the coefficients. It is important to
mention that every  $a_k$ is annihilated
by ${\cal O}$; this was proven in \CachazoDR. 

The way to deal with this problem is to realize that for any two
functions $G_1$ and $G_2$ satisfying ${\cal O}^2(G_k)=0$, the
following combination
\eqn\comb{ H(G_1, G_2) = {\cal O}(G_1)G_2  - {\cal O}(G_2)G_1}
is annihilated by ${\cal O}$.  In the calculations we have done, the factors $G_k$ arising from the cut integrals all satisfy ${\cal O}^2(G_k)=0$, and we believe that this property is satisfied generally.

Therefore, any rational function with both factors in the
denominator ``splits" as follows
\eqn\split{ {P\over QG_1 G_2 \prod'_{k} G_k} = {P\over Q
\prod'_{k}G_k} \left( { {\cal O}(G_1) \over G_1} - {{\cal
O}(G_2)\over G_2} \right)\times {1\over H(G_1,G_2)}, }
where $\prod'$ means a product not including $G_1$ or $G_2$.

It is clear that this procedure can be repeated as many times as
necessary until the original rational function \ratt\ is written
in the form
\eqn\heli{ {P\over Q\prod_{k} G_k} = \sum_{k} {P_k\over Q_k}
{{\cal O}(G_k)\over G_k}.}
This formula is very similar to what we want \scheme. However, the
procedure just described only guarantees that ${\cal O}Q_k = 0$,
but in general the same is not true of $P_k$. Recall that the
coefficients $a_k$, which we are after,  are annihilated by ${\cal
O}$.

The way out of this problem is to realize that near a kinematical
region\foot{We thank Oleg Lunin for suggesting to look at this
particular regime.} where a given $G_l=0$ we should find
\eqn\limit{ {P_l\over Q_l} \to a_l.}
Since $P_l$ is a polynomial, this implies that $P_l$ admits an
expansion of the form
\eqn\expa{ P_l = Q_l a_l + \sum_{m=1}^{\infty} h_m (G_l)^m, }
where most terms in the sum are zero because $P_l$ has a finite
degree. Note that $P_l - Q_l a_l$ is a polynomial divisible by
$G_l$. Therefore it can be written as $P_l - Q_l a_l = G_l X_l$,
where $X_l$ is some polynomial.  We think of this as a kind of ``polynomial division.''

The decomposition of $P_l$ in the form \expa\ is easily done by
introducing coordinates where $G_l$ is one variable and all other
variables are kinematical invariants which are annihilated by
${\cal O}$. This guarantees that $Q_l a_l$ is annihilated by
${\cal O}$, as it should be.

After this is done for each $P_k$ in \heli, we are left with
\eqn\lefty{{\cal O}C_{i,i+1,\ldots ,j-1,j} =  \sum_{k} a_k {{\cal
O}(G_k)\over G_k} + \sum_k {X_k\over Q_k}{\cal O}(G_k).}
Comparing \lefty\ to \scheme\ we find that a miraculous
cancellation must take place, namely
\eqn\jima{\sum_k {X_k\over Q_k}{\cal O}(G_k) = 0. }
Indeed, we find this cancellation in all the cuts considered in
the next section.

In practice, the splitting procedure is done most efficiently as follows.  
The operation performed in \split\ splits the rational function into two terms, such that $G_1$ appears only in the denominator of one term and $G_2$ appears only in the denominator of the other.
To determine the coefficient ${P_1 / Q_1}$ in \heli, all we need is to isolate the factor $G_1$ from all other factors $G_k$, one factor at a time.  That is, if $k$ runs from 1 to $r$, we apply the operation \split\ $r-1$ times, and each time, we keep only the term with $G_1$ remaining in the denominator.  The result is that
\eqn\practical{ {P_1\over Q_1}{{\cal O}(G_1)\over G_1}
= {\cal O} C_{i,i+1,\ldots ,j-1,j}\times \prod_{k=2}^r { {\cal O}(G_1) G_k\over H(G_1,G_k) }.
}
Thus, computing all $r$ coefficients (before performing the polynomial division) requires a total of only $r(r-1)$ operations.  The point is that it is most efficient to obtain first the coefficient of one factor, dropping terms that do not contain it, and then start over for the next factor.

\subsec{Collinear Operators}

The question is now how to construct differential operators that
produce rational functions when acting on the cut integral \cutIn.
In \CachazoDR, a simple prescription was given. Consider any
operator $F_{ijk}$ that tests whether gluons $i$, $j$, and  $k$
are localized on a line in twistor space. (These operators were
originally introduced in section 3 of \WittenNN. For a short
review see section 2 of \CachazoDR.)

These are defined in the spinor-helicity formalism of
\refs{\berends,\xu,\gunion} as follows:
\eqn\expis{ F_{ijk;\dot a} = \vev{i~j}{\del \over \del
\tilde\lambda_k^{\dot a}} + \vev{k~i}{\del \over \del
\tilde\lambda_j^{\dot a}} + \vev{j~k}{\del \over \del
\tilde\lambda_i^{\dot a}},}
where $\dot a$ is a negative chirality spinor index. Therefore
$F_{ijk;\dot a}$ is a spinor-valued differential operator.

In the following, it will be convenient to introduce a fixed,
arbitrary, negative-chirality spinor $\eta^{\dot a}$ and consider
\eqn\prox{ [F_{ijk}, \eta] = \epsilon^{\dot a\dot b}\eta_{\dot
a}F_{ijk;\dot b}. }
Note that the brackets in \prox\ are meant to indicate the inner
product of two negative chirality spinors and not the commutator
of operators.

Naively, any operator that tests the collinearity in twistor space
of three gluons in the MHV amplitude of \cutIn\ annihilates the
cut integral. This is because tree-level MHV amplitudes are
localized on a line \WittenNN. However, it was found in
\CachazoBY\ that the cut integral has a holomorphic anomaly that
spoils this result. Instead, the collinear operator produces a
delta function that localizes the integral completely when
$\ell_1$ or $\ell_2$ participates in it. Therefore, it produces a
rational function.

Going back to the particular cut integral \cutIn, it turns out
that the only collinear operators that localize the integral are
those of the form $[F_{ikl},\eta]$ and $[F_{klj},\eta]$, where
$k,l$ are any gluons participating on the left side of the
cut.\foot{Of course, if $k$ or $l$ is equal to $i$ ($j$) then the
operator $[F_{ikl},\eta]$ ( $[F_{klj},\eta]$) vanishes
trivially.}

Consider for example the action of the collinear operator
$[F_{i,i+1,i+2},\eta]$ on the cut integral $C_{i,i+1,\ldots ,j}$.

In order to describe the rational function very explicitly, we have
to exhibit the explicit dependence on the spinors
$\lambda_{\ell_1}$ and $\tilde\lambda_{\ell_1}$ of the tree-level
amplitude on the right in \cutIn:
\eqn\explix{A^{\rm tree}(\ell_2,j+1,j+2,\ldots ,i-1,\ell_1) =
A^{\rm tree}(\ell_2,j+1,j+2,\ldots ,i-1,\{
\lambda_{\ell_1},\tilde\lambda_{\ell_1}\} ). }

Now we are ready to write the action of the operator \CachazoDR\foot{A similar formula was obtained for MHV one-loop amplitudes in \Bena.}:
\eqn\excutre{\eqalign{ & [F_{i,i+1,i+2},\eta]C_{i,i+1,\ldots
,j-1,j} = \cr & {t\over (2p_i\cdot P_L)}{\vev{k~m}^4\over
\vev{i~i+1}\ldots \vev{j-1~j}} {\vev{i+1~i+2}[i~\eta]\over
\vev{\ell_2~i}\vev{j~\ell_2}}A^{\rm tree}(\ell_2,j+1,j+2,\ldots
,i-1,\{ \lambda_i , t\tilde\lambda_i \} ),}}
with
\eqn\husa{ \ell_2 = P_L - t p_i, \qquad t = { P_L^2\over
(2p_i\cdot P_L)}, \qquad P_L = p_i + p_{i+1}+\ldots + p_j.}

All we need is to put the explicit form of the tree-level amplitude on
the right, make the substitutions and apply the procedure
described above with the generic operator ${\cal O}$ replaced by
$[F_{i,i+1,i+2},\eta]$.   To  reconstruct the whole amplitude, we need to know that the coefficient of every scalar box function in \gene\ can be calculated from one of the cuts.  This is proven in Appendix B.

To illustrate this technique, we compute the full next-to-MHV leading-color ${\cal N}=4$
seven-gluon amplitude $A_{7:1}(1^-,2^-,3^-,4^+,5^+,6^+,7^+)$.

%%%%%%%%%%%%%%%%%%%%%%%%%%%%%%%%%%%%%%%%%%%%
\newsec{Computation of $A_{7:1}(1^-,2^-,3^-,4^+,5^+,6^+,7^+)$}

In this paper, we compute the seven-gluon amplitude with the particular helicity configuration $(---++++)$.   All other helicity configurations of seven gluons could be computed in just the same way, with no new ingredients.

The amplitude $A_{7:1}(1^-,2^-,3^-,4^+,5^+,6^+,7^+)$ is expressed
in terms of thirty-five box functions.  We abbreviate the indices
on the coefficients of \gene\ for simplicity.
\eqn\bcddg{\eqalign{
A_{7:1}(1^-,2^-,3^-,4^+,5^+,6^+,7^+) & =  \cr
= \sum_{i=1}^7 \left( b_i F^{1m}_{7:i} + \right. & \left. c_i
F^{2m~e}_{7:2;i}+d_{2,i}F^{2m~h}_{7:2;i} +d_{3,i}F^{2m~h}_{7:3;i}+
g_i F^{3m}_{7:2:2;i}\right). }}
Ten of these were already computed in \CachazoDR~from the
$C_{123}$ cut, namely\foot{We conjugate the coefficients of
\CachazoDR, which were derived for  the seven-gluon one-loop amplitude
$A_{7:1}(1^+,2^+,3^+,4^-,5^-,6^-,7^-)$ with the opposite helicity
assignments.}
\eqn\ekibu{\eqalign{ &b_4=c_5=d_{2,2}=d_{3,5}= { (t_1^{[3]})^3
\over
[1~2][2~3]\vev{4~5}\vev{5~6}\vev{6~7}\gb{4|2+3|1}\gb{7|1+2|3}},
\cr & c_1=c_2=d_{2,6}=d_{3,1}=g_2=g_4=0. }}
Here we have defined $$\gb{i|j_r+j_{r+1}+\cdots+j_s|k} \equiv
\vev{i~j_r}[j_r~k]+\vev{i~j_{r+1}}[j_{r+1}~k]+\cdots+\vev{i~j_s}[j_s~k].$$

We apply our reduction technique first by applying
$[F_{456},\eta]$ on the cut $C_{456}$.  This yields five more coefficients.  Next, we apply
$[F_{712},\eta]$ on the cut $C_{712}$.  This calculation is
slightly more involved, because here it is possible for fermions
and scalars to circulate in the loop.  We find seven more coefficients.
We can obtain corresponding
results for the cuts $C_{567}$ and $C_{234}$ simply by permuting
the labels, for nine new coefficients.  At this point
we have found thirty-one of the thirty-five coefficients.  
The remaining four are easily determined by the known infrared behavior of the amplitude.

% % % % % % % % % %
\subsec{The Cut $C_{456}$}

The cut $C_{456}$ is given by \eqn\wholeffscut{\eqalign{C_{456} &=
{\rm Im}|_{t_4^{[3]}>0} ~ \left( c_{1} F^{2m~e}_{7:2;1} + d_{2,2}
F^{2m~h}_{7:2;2} +d_{3,4} F^{2m~h}_{7:3;4}+ \right. \cr & \left.
 + b_7 F^{1m}_{7:7} + c_4 F^{2m~e}_{7:2;4} + c_5 F^{2m~e}_{7:2;5}
+ d_{2,5} F^{2m~h}_{7:2;5}  +d_{3,1} F^{2m~h}_{7:3;1} + g_{5}
F^{3m}_{7:2:2;5} + g_{7} F^{3m}_{7:2:2;7} \right) }}
or by the cut integral
\eqn\domas{  C_{456} =  \int d\mu~~ A^{\rm tree}((-\ell_1)^-,
4^+, 5^+, 6^+, (-\ell_2)^-) A^{\rm
tree}(\ell_1^+,\ell_2^+,7^+,1^-,2^-,3^-).}
Note that in this case only gluons can run in the loop and that
the five-gluon tree-level amplitude is an MHV amplitude. According
to the general discussion of section 2, we should consider the
action of the collinear operator $[F_{456}, \eta]$ on both
\wholeffscut\ and \domas.

The first step is to calculate the action of the collinear
operator $[F_{456},\eta]$ on $C_{456}$ given by \wholeffscut. Note
that the three box functions in the top line of \wholeffscut\ are
annihilated by the operator, so we cannot calculate those
coefficients directly using this operator. Let us list the
imaginary parts of the relevant scalar box functions in the
kinematical regime where $t_4^{[3]}>0$ and all other invariants
are negative.\foot{In these expressions we suppress an 
overall factor of $\pi$.}
\eqn\imag{ \eqalign{ {\rm Im}|_{t_4^{[3]}>0} F^{1m}_{7;7} =&~
-\ln\left( 1- {t_4^{[3]}\over t_4^{[2]}} \right) - \ln\left( 1-
{t_4^{[3]}\over t_5^{[2]}} \right) \cr  {\rm
Im}|_{t_4^{[3]}>0}F^{2m~e}_{7:2;5} =&~ \ln\left( 1-
{t_5^{[2]}\over t_4^{[3]}} \right) - \ln\left( 1-
{t_5^{[2]}t_1^{[3]}\over t_4^{[3]}t_5^{[3]}} \right)+\ln\left(
-{t_4^{[3]}\over t_5^{[3]}} \right) + \ldots \cr  {\rm
Im}|_{t_4^{[3]}>0}F^{2m~e}_{7:2;4} =&~ \ln\left( 1-
{t_4^{[2]}\over t_4^{[3]}} \right) - \ln\left( 1-
{t_4^{[2]}t_7^{[3]}\over t_3^{[3]}t_4^{[3]}} \right)+\ln\left(
-{t_4^{[3]}\over t_3^{[3]}} \right) + \ldots  \cr {\rm
Im}|_{t_4^{[3]}>0}F^{2m~h}_{7:2;5} =&~ \ln\left( -{t_4^{[3]}\over
t_3^{[2]}} \right) + \ln\left( 1- {t_5^{[2]}\over t_4^{[3]}}
\right) + \ldots \cr {\rm Im}|_{t_4^{[3]}>0}F^{2m~h}_{7:3;1} =&~
\ln\left( -{t_4^{[3]}\over t_6^{[2]}} \right) + \ln\left( 1-
{t_4^{[2]}\over t_4^{[3]}} \right) + \ldots  \cr {\rm
Im}|_{t_4^{[3]}>0}F^{3m}_{7:2:2;5} =&~ \ln\left( -{t_4^{[3]}\over
t_5^{[4]}} \right) + \ln\left( 1- {t_5^{[2]}\over t_4^{[3]}}
\right) - \ln\left( 1- {t_5^{[2]}t_2^{[2]}\over
t_4^{[3]}t_5^{[4]}} \right) + \ldots \cr  {\rm
Im}|_{t_4^{[3]}>0}F^{3m}_{7:2:2;7} =&~ \ln\left( -{t_4^{[3]}\over
t_6^{[3]}} \right) + \ln\left( 1- {t_4^{[2]}\over t_4^{[3]}}
\right) - \ln\left( 1- {t_7^{[2]}t_4^{[2]}\over
t_6^{[3]}t_4^{[3]}} \right) + \ldots }}
The ellipses represent terms that are annihilated by the
collinear operator $[F_{456},\eta]$. In other words, the terms represented
by ellipses depend on $p_4$, $p_5$, and $p_6$ only through the
combination $p_4+p_5+p_6$.

Now we can compute the action of the collinear operator on
$C_{456}$ given by the imaginary part of the amplitude
\wholeffscut. Here we denote $[F_{456},\eta]$ by ${\cal O}$ in
order to make contact with the general discussion of section 2 and
to avoid cluttering the equations.
\eqn\fenix{\eqalign{ {\cal O}C_{456} =&~ b_7{{\cal
O}(t_4^{[2]}t_5^{[2]})\over t_4^{[2]}t_5^{[2]}} - c_5 {{\cal
O}(t_5^{[2]}t_1^{[3]}-t_4^{[3]}t_5^{[3]})\over
t_5^{[2]}t_1^{[3]}-t_4^{[3]}t_5^{[3]}} - c_4 {{\cal
O}(t_4^{[2]}t_7^{[3]}-t_3^{[3]}t_4^{[3]})\over
t_4^{[2]}t_7^{[3]}-t_3^{[3]}t_4^{[3]}} + d_{2,5}{{\cal
O}(t_3^{[2]})\over t_3^{[2]}} \cr & + d_{3,1}{{\cal
O}(t_6^{[2]})\over t_6^{[2]}} - g_5 {{\cal
O}(t_5^{[2]}t_2^{[2]}-t_4^{[3]}t_5^{[4]})\over
t_5^{[2]}t_2^{[2]}-t_4^{[3]}t_5^{[4]}} - g_7 {{\cal
O}(t_7^{[2]}t_4^{[2]}-t_6^{[3]}t_4^{[3]})\over
t_7^{[2]}t_4^{[2]}-t_6^{[3]}t_4^{[3]}} \cr  & +
(-b_7+c_4+d_{3,1}+g_7){{\cal O}(t_4^{[3]} - t_4^{[2]}) \over
t_4^{[3]} - t_4^{[2]}} + (-b_7+c_5+d_{2,5}+g_5) {{\cal
O}(t_4^{[3]} - t_5^{[2]}) \over t_4^{[3]} - t_5^{[2]}}.}}
We have written in the first two lines the contributions from the
poles that uniquely identify a given scalar box function. This is
manifest from the fact that only one coefficient appears in front
of each of them. On the other hand, the poles in the third line
are common to several box functions and so their coefficients are
linear combinations of the scalar box function coefficients.

We now turn to the computation of the action of the collinear
operator on the cut integral representation of $C_{456}$. The cut
integral \didi\ is written as
\eqn\didiffs{  C_{456} =  \int d\mu~~  {\vev{\ell_2~\ell_1}^3
\over \vev{\ell_1~4}\vev{4~5}\vev{5~6}\vev{6~\ell_2}} A_6^{\rm
tree}(\ell_1^+,\ell_2^+,7^+,1^-,2^-,3^-),}
where for the tree-level six-gluon amplitude we use a result from
\refs{\mangpxu,\mangparke}:
\eqn\mpxsix{\eqalign{A_6^{\rm tree}(1^-,2^-,3^-,\ell_1^+,\ell_2^+,7^+) &=
\left[ {\beta^2 \over t_{\ell_2 71}s_{\ell_2 7}s_{71}s_{23}s_{3\ell_1}} +
{\gamma^2 \over t_{712}s_{71}s_{12}s_{3\ell_1}s_{\ell_1 \ell_2}} \right. \cr
&~~~~~~~~~~~~~+ \left.{\beta \gamma
t_{\ell_1 \ell_2 7} \over s_{\ell_1 \ell_2}s_{\ell_2 7}s_{71}s_{12}s_{23}s_{3\ell_1}} \right], \cr
\beta &= [\ell_2~7]\vev{2~3}\gb{1|\ell_2+7|\ell_1}, \cr \gamma &=
[\ell_1~\ell_2]\vev{1~2}\gb{3|\ell_1+\ell_2|7}, \cr
s_{ij}&=\vev{i~j}[i~j], \cr
t_{ijk}&=\vev{i~j}[i~j]+\vev{i~k}[i~k]+\vev{j~k}[j~k]
.}}
This amplitude could be written in terms of the MHV diagrams of \CachazoKJ.  In this case, the formula in \mpxsix\ is simpler, but for more gluons we expect the MHV diagrams  to be most efficient.

The  integral \didiffs\ is of the form analyzed in section 2. Here we want
to compute the action of $[F_{456},\eta]$ to $C_{456}$. We can
simply apply the general formula \excutre\ to get the result. Note
that \excutre\ is the result of the action of the operator on a
single pole. In the case at hand, the operator $[F_{456},\eta]$
acts nontrivially on two poles, namely $1/\vev{\ell_1~4}$ and
$1/\vev{6~\ell_2}$. This only means that we have to apply
\excutre\ twice and add the results.

Consider first the action on the pole $1/\vev{\ell_1~4}$.
We find
\eqn\spofffs{\eqalign{ ([F_{456},\eta] C_{456})^{\rm first} &= { [4~\eta] (t_4^{[3]})^2 \over \vev{5~6} t_4^{[2]}}
\left[ {\beta_1^2 \over (t_5^{[2]} t_2^{[2]}-t_4^{[3]}t_2^{[3]})
(t_5^{[2]} t_4^{[4]}-t_4^{[3]}t_5^{[3]})
t_7^{[2]}t_2^{[2]}t_3^{[2]}}\right.  \cr   &~~+\left. {\gamma_1^2 \over
t_7^{[3]}t_7^{[2]}t_1^{[2]} t_3^{[2]} t_4^{[3]}} + {\beta_1 \gamma_1
t_4^{[4]} \over t_4^{[3]} (t_5^{[2]}
t_4^{[4]}-t_4^{[3]}t_5^{[3]} ) t_7^{[2]}t_1^{[2]}t_2^{[2]} t_3^{[2]}} \right], \cr \beta_1 &=
-\vev{2~3}\gb{4|5+6|7} \gb{1|5+6|4}, \cr \gamma_1 &=
\vev{1~2}\gb{3|4+5+6|7}. }}
We identify the four poles important to this cut as those factors
in the denominator not annihilated by $[F_{456},\eta]$.  These are
$t_4^{[2]},t_3^{[2]},(t_5^{[2]}t_4^{[4]}-t_4^{[3]}t_5^{[3]}),(t_5^{[2]} t_2^{[2]}-t_4^{[3]}t_2^{[3]})$, which appear
respectively (and uniquely) in the box functions
$F^{1m}_{7:7},F^{2m~h}_{7:2;5},F^{2m~e}_{7:2;5},F^{3m}_{7:2:2;5},$.
These four poles are the $G_k$ of the previous section.  Now we
apply our procedure to separate the cut into simple fractions. For
example, to isolate the particular pole
$G_0=(t_5^{[2]} t_2^{[2]}-t_4^{[3]}t_2^{[3]})$, we evaluate
\eqn\splitsville{([F_{456},\eta] C_{456})^{\rm first} \times
 \left( { t_4^{[2]}~{\cal
O}(G_0)\over H(G_0,t_4^{[2]})} \right) \left( { t_3^{[2]}~{\cal O}(G_0)
\over H(G_0,t_3^{[2]})} \right) \left( {
(t_5^{[2]} t_4^{[4]}-t_4^{[3]}t_5^{[3]})~{\cal O}(G_0) \over
H(G_0,(t_5^{[2]} t_4^{[4]}-t_4^{[3]}t_5^{[3]}))} \right).} 
Perform
the ``polynomial division'' of section 2 on the numerator to separate the ``extra''
part proportional to $G_0.$ It simplifies computations to
perform the operations \splitsville\ on each term of \spofffs\ separately, for
only the poles that appear in that term.  As long as the procedure
is consistent for all poles in each term, it is valid. After all,
we are multiplying by factors that appear in pairs that sum to 1.
As long as the arguments $G_k$ of $H$ satisfy ${\cal O}^2(G_k) =
0$, we can use any ones we like.

The first check that our procedure is working is that \jima\ is satisfied: 
the ``extra'' parts from each of the four poles sum to zero.

The remainder of \splitsville\ is found to be of the form
\eqn\triumph{-c_5 {{\cal O}(G_0)\over G_0}.} We now have our first
coefficient, $c_5,$ and our second consistency check, because its
conjugate was already computed in \CachazoDR.  Indeed, our result
agrees:\eqn\gotcfive{ c_5 = {(t_1^{[3]})^3 \over [1~2]
[2~3]\vev{4~5}\vev{5~6}\vev{6~7}\gb{4|2+3|1}\gb{7|1+2|3}}.}
The other three coefficients calculated from 
$([F_{456},\eta] C_{456})^{\rm first}$
are $b_7,d_{2,5},$ and $g_5.$ \eqn\gotdtwofive{\eqalign{ d_{2,5}
&= {\vev{1~2}^3 (t_4^{[3]})^3 \over \vev{4~5}\vev{5~6}\vev{7~1}
t_7^{[3]}\gb{7|1+2|3}\gb{6|4+5|3} (\vev{4~2} t_4^{[3]} +\vev{2~3}
\gb{4|5+6|3})}, \cr g_{5} &= {\vev{2~3}^3 \gb{4|5+6|7}^3 \over
\vev{3~4}\vev{4~5}\vev{5~6} [7~1] \gb{4|2+3|1} (\vev{4~2}
t_4^{[3]} + \vev{2~3} \gb{4|5+6|3}) (\vev{5~6} \gb{4|2+3|5} -
\vev{4~6} t_2^{[3]})}. }}
The expression for $b_7$ was found, but by itself is too
complicated to write here.  We will have more to say on this
presently.

The action of ${\cal O}$ on the second pole $1/\vev{6~\ell_2}$
similarly yields four coefficients: \eqn{\gotcfour}{\eqalign{
c_{4} &= { \gb{3|1+2|7}^3 \over [7~1] [1~2]
\vev{3~4}\vev{4~5}\vev{5~6} t_7^{[3]}\gb{6|7+1|2}}, \cr d_{3,1} &=
0, \cr g_{7} &= {(\vev{6~1} t_4^{[3]} - \vev{7~1} \gb{6|4+5|7})^3
\over [2~3]\vev{4~5}\vev{5~6}\vev{6~7}\vev{7~1}
\gb{6|4+5|3}\gb{6|7+1|2} (\vev{4~6} t_6^{[3]} - \vev{4~5}
\gb{6|7+1|5})}.}}
The coefficient $b_7$ appears here too and agrees with the
expression computed from the other term.  Moreover, we can check
two more relations among these coefficients.  The box functions
participating in this cut have some poles that do not appear in
the integral.  These are $(t_4^{[3]}-t_4^{[2]})$ and
$(t_4^{[3]}-t_5^{[2]})$. Eq. \fenix\ then implies the two
relations $-b_7+c_4+d_{3,1}+g_7=0$ and $-b_7+c_5+d_{2,5}+g_5=0$.
We have checked that our coefficients do indeed satisfy these
relations. In section 4, we will use the first relation to list
$b_7$ in terms of $c_4$ and $g_7$, but we must stress that we have
computed it independently.

To summarize, the cut $C_{456}$ involves the ten coefficients seen
in \wholeffscut. We have computed the seven that appear on the
second line.  Two of the coefficients of the first line are known
from \bcddg:  $c_1=d_{2,2}=0$.  The last coefficient, $d_{3,4}$,
will show up in the cut we compute next.

%%%%%%%%%%%%%%%%%%%%%%
\subsec{The Cut $C_{712}$}
%%%%%%%%%%%%%%%%%%%%%%

The cut $C_{712}$ is given by \eqn\wholesotcut{\eqalign{C_{712} &=
{\rm Im}|_{t_7^{[3]}>0} ~ \left( c_{4} F^{2m~e}_{7:2;4} + d_{2,5}
F^{2m~h}_{7:2;5} +d_{3,7} F^{2m~h}_{7:3;7}+ \right. \cr & \left.
 + b_3 F^{1m}_{7:3} + c_1 F^{2m~e}_{7:2;1} + c_7 F^{2m~e}_{7:2;7}
+ d_{2,1} F^{2m~h}_{7:2;1}  +d_{3,4} F^{2m~h}_{7:3;4} + g_{1}
F^{3m}_{7:2:2;1} + g_{3} F^{3m}_{7:2:2;3} \right). }}
For this cut, there are three possible helicity assignments for
$\ell_1, \ell_2$. If we denote the helicity of $(\ell_1, \ell_2)$
by the assignment on the  amplitude $A^{\rm tree}(\ell_1,7^+,1^-,2^-, \ell_2)$,
these three cases are:  (a) $(\ell_1, \ell_2)=(+,-)$; (b)
$(\ell_1, \ell_2)=(+,-)$; (c) $(\ell_1, \ell_2)=(+,+)$. Notice
that the assignment $(\ell_1, \ell_2)=(-,-)$ does not contribute,
because the amplitude $A^{\rm tree}(\ell_1^-,7^+,1^-,2^-, \ell_2^-)$ 
vanishes.

Now let us discuss these three assignments. For cases (a) and (b),
the particle circulating in the loop can be a gluon, fermion or
complex scalar of the ${\cal N}=4$ multiplet. Thus the expression
will be\foot{We use $n$ for generality. In our particular example,
$n=7$.} \eqn\Cab{\eqalign{ C_{n12}^{(a/b)} & =
\int d\mu~ A^{{\rm tree},V} ((-\ell_1)^{\pm}, n^+,1^-, 2^-,
(-\ell_2)^{\mp}) A^{{\rm tree},V} (\ell_2^{\pm}, 3^-, 4^+,..., (n-1)^+,
\ell_1^{\mp}) \cr & + (-4)  \int d\mu~ A^{{\rm tree},F}
((-\ell_1)^{\pm}, n^+,1^-, 2^-, (-\ell_2)^{\mp}) A^{{\rm tree},F}
(\ell_2^{\pm}, 3^-, 4^+,..., (n-1)^+, \ell_1^{\mp}) \cr & + (+3)
\int d\mu~ A^{{\rm tree},S} ((-\ell_1)^{\pm}, n^+,1^-, 2^-,
(-\ell_2)^{\mp}) A^{{\rm tree},S} (\ell_2^{\pm}, 3^-, 4^+,..., (n-1)^+,
\ell_1^{\mp}), }} where $(-4)$ counts the four fermions and $(+3)$
counts the three complex scalars in the ${\cal N}=4$ multiplet.
The supersymmetric Ward identity relates fermion and scalar MHV
amplitudes to gluon MHV amplitudes by \refs{\grisaru,\mangparke} 
\eqn\MHVF{\eqalign{ A(
F_1^-, g_2^+,.., g_j^-,..., F_n^+) & =  { \vev{ j ~ n}\over \vev{j
~  1}} A^{\rm MHV}( g_1^-, g_2^+,..., g_j^-,..., g_n^+), \cr A( S_1^-,
g_2^+,.., g_j^-,..., S_n^+) & = { \vev{ j  ~ n}^2\over \vev{j ~
1}^2} A^{\rm MHV}( g_1^-, g_2^+,..., g_j^-,..., g_n^+). }} We need to
be careful about the ordering when $\ell_1, \ell_2$ are fermions.
They should be ordered according to \Cab. If $F^+$ and  $F^-$ 
exchange positions in \MHVF, there is an extra
$(-)$ sign. Having taken care of the ${\cal N}=4$ multiplet we
have\foot{The $(-)^5$ sign comes from the left hand part since it
is $\overline{\rm MHV}$. The rule to go from MHV to $\overline{\rm
MHV}$ is to exchange $\vev{~}\leftrightarrow [~]$ and multiply by
$(-)^n$.}
\eqn\mtare{\eqalign{
&C_{n12}^{(a)+(b)}= \cr &{ (-)^5\over
[n~1][1~2]\vev{3~4}\vev{4~5}...\vev{(n-2)(n-1)} } \int d\mu { \rho^2
[\ell_1 ~n]^2 [\ell_2 ~n]^2 \vev{ 3 ~\ell_1}^2 \vev{3 ~\ell_2}^2 \over
[\ell_1 ~n] [2 ~\ell_2] [\ell_2 ~\ell_1] \vev{\ell_2 ~3} \vev{(n-1)
\ell_1}\vev{\ell_1 ~\ell_2}, }}}

where \eqn\RHO{\eqalign{ & \rho^2 = 
\cr
 & \left( {\vev{3 ~\ell_2}^2
[\ell_2 ~n]^2 \over\vev{3 ~\ell_1}^2 [\ell_1~n]^2}
 \right)^{2}
+4 \left( {\vev{3 ~\ell_2}^2 [\ell_2 ~n]^2 \over\vev{3 ~\ell_1}^2
[\ell_1 ~n]^2}
 \right) +6 +4 \left( {\vev{3 ~\ell_2}^2 [\ell_2 ~n]^2 \over\vev{3 ~\ell_1}^2
[\ell_1 ~n]^2}
 \right)^{-1}+
\left( {\vev{3 ~\ell_2}^2 [\ell_2 ~n]^2 \over\vev{3 ~\ell_1}^2
[\ell_1 ~n]^2}
 \right)^{-2} 
\cr
 & = { \gb{ 3|
(n+1+2)|n}^4 \over [\ell_1 ~n]^2 [\ell_2 ~n]^2 \vev{ 3 ~\ell_1}^2
\vev{3 ~\ell_2}^2} .
}}
Making the substitution for $\rho^2,$ we get 
\eqn\CNAB{\eqalign{
& C_{n12}^{(a)+(b)}  = \cr &{\gb{ 3| (n+1+2)|n}^4 \over (t_n^{[3]})^4} {
(-) (-)^5
 (t_n^{[3]})^3 \over [n~1][1~2]\vev{3~4}\vev{4~5}...\vev{(n-2)~(n-1)} }
\int d\mu {1\over [\ell_1 ~n] [2 ~\ell_2] \vev{\ell_2 ~3} \vev{(n-1)
~\ell_1}} \cr & ={\gb{ 3| (n+1+2)|n}^4 \over (t_n^{[3]})^4}
[C_{123}^{\dagger}]|_{j \rightarrow j-1}. }} Using the result of
\CachazoDR\ for $C_{123}^{\dagger}$, we can read out the
contribution of the $(a)+(b)$ part to the following coefficients
(with $n=7$): 
\eqn\AandB{ b_3^{(a)+(b)} =
c_{4}^{(a)+(b)}=d_{2,1}^{(a)+(b)}=d_{3,4}^{(a)+(b)} ={\gb{ 3|
(1+2)|7}^3 \over (t_7^{[3]}) [7~1][1~2]\vev{3~4}\vev{4~5} \vev{5~6}
\gb{ 6| 7+1| 2}}. }

Now we discuss the assignment (c) given by\foot{Relative to
assignments (a) and (b), there is an extra $(-)$ sign. The reason is
that for the assignment (c) the left hand side is MHV already, so
we do not have the $(-)^5$ factor here.}
$$
C_{712}^{(c)}= \int d\mu~ A^{\rm tree}((-\ell_1)^+, 7^+, 1^-, 2^-,
(-\ell_2)^+) A^{\rm tree}(4^+, 5^+, 6^+, \ell_1^-, \ell_2^-, 3^-)
$$
Notice that for the assignment (c), only gluons can propagate
along internal lines.  The first factor is again an MHV amplitude,
so we can directly apply the general method of section 2. The
second factor has the same helicity structure $(+++---)$ that we
saw in the previous cut, making this computation very similar to the
the previous one. The collinear operator acts on two poles, namely
$1/\vev{7~\ell_1}$ and $1/\vev{2~\ell_2}$. Each of the terms thus
obtained involves four unique poles of the scalar box functions in
\wholesotcut.  We apply the reduction procedure to produce the
following coefficients (after again confirming \jima, that all the
``extra'' pieces sum to zero). 
\eqn\Cpart{\eqalign{ d_{2,1}^{(c)}
& = - { \gb{ 3| 4+5|6}^3 \vev{1~2}^3 \over \vev{7~1}\vev{3~4}\vev{4~5}
t_{3}^{[3]} t_{7}^{[3]} \gb{ 2|7+1|6} ( \vev{6~5} \gb{7|1+2|
6}-\vev{7~5} t_{7}^{[3]})}, \cr 
g_1^{(c)} & = { \vev{1~2}^3 \gb{7 |
5+6|4}^3 \over \vev{5~6} \vev{6~7} \vev{7~1} [3~4] \gb{7| 1+2| 3} (
\vev{7~2} t_5^{[3]} +\vev{2~1} \gb{7|5+6|1}) (\vev{6~5}  \gb{7| 1+2|
6}-\vev{7~5} t_7^{[3]})},\cr c_7^{(c)} & = - { \vev{1~2}^3 [5~4]^3
\over t_{3}^{[3]} \vev{6~7} \vev{7~1} [3~4] \gb{2| 3+4 |5} \gb{ 6|
4+5 | 3}},\cr g_3^{(c)} & = - { \vev{1~2}^3 \vev{2~3}^3 [5~6]^3 \over
\vev{7~1} \vev{3~4} \gb{2 | 3+4| 5} \gb{ 2| 7+1|6} ( \vev{7~1} \gb{
2| 3+4 |1} - t_{2}^{[3]} \vev{7~2}) ( t_{3}^{[4]}\vev{2~4} -\vev{3~4}
\gb{ 2|7+1| 3})}, \cr 
d_{3,4}^{(c)} & = - { \vev{1~2}^3 (t_{4}^{[3]})^3 \over
\vev{4~5} \vev{5~6} \vev{7~1} t_{7}^{[3]} \gb{ 7| 1+2 |3} \gb{6|
4+5|3} (  t_{3}^{[4]} \vev{2~4} -\vev{3~4} \gb{ 2|7+1| 3})}, }} 
and a
complicated expression for $b_3^{(c)}$.  Here, the analog of \fenix\ from the previous case is the same equation but with all indices shifted by $+3$.  This is because box functions are oblivious to helicity.  As before, there are two relations derived from the poles present in the box functions that do not appear in the cut integral.  They are $-b_3+c_7+d_{3,4}=0$ and $-b_3+c_1+d_{2,1}+g_1=0$.
We have confirmed that both of these relations are satisfied.

We now have explicit expressions for nine of the ten coefficients appearing in \wholesotcut.  The seven coefficients appearing in the second line, have just been computed by our reduction method, and $c_4$ and $d_{2,5}$ were evaluated in the previous cut.  (We did find a contribution to $c_4$ again in \AandB.  But remember that the operator $[F_{712},\eta]$ gives no information about the coefficients in the first line of \wholesotcut, because those box functions are annihilated.  Therefore $c_4^{(c)}$ is undetermined, and we must take the result for $c_4$ from the previous cut.)  It is possible to find the single remaining coefficient, $d_{3,7}$, by imposing the finiteness of this cut.  All cuts in three-particle channels are finite.  This condition is discussed and derived in Appendix A.
\eqn\llpub{
d_{3,7} = -2
b_3+ 2c_7-2c_4+2d_{3,4}-d_{2,5}+2d_{2,1}+ g_3+ g_1.}
Incidentally, now that we have computed $d_{3,4}$ explicitly, it is possible to test the finiteness of the cut $C_{456}$ as a consistency check.  
This condition, derived similarly, is
\eqn\llpubong{
0=-b_7-c_1+c_4+c_5-\half d_{2,2} + d_{2,5}+d_{3,1}-\half d_{3,4} +
\half g_5 + \half g_7.}

%%%%%%%%%%%%%%%%%%%%%%%%%
\subsec{The Cuts $C_{567}$ And $C_{234}$: Reflection Of Indices}
%%%%%%%%%%%%%%%%%%%%%%%%%

Knowing  the contributions from the cuts $C_{456}$ and $C_{712}$,
 we can use  reflection symmetry of the indices
to get the  contributions from cuts $C_{567}$ and $C_{234}$
without further calculations.  Under the reflection of indices
\eqn\REF{ \sigma : \quad 1\leftrightarrow 3,~~~4\leftrightarrow
7,~~~5\leftrightarrow 6,~~~ \ell_1\leftrightarrow  \ell_2, } every
possible helicity assignment of $\ell_1, \ell_2$
 of, for example, cut $C_{456}$ is mapped to
a unique corresponding helicity assignment of $\ell_1, \ell_2$
 of cut $C_{567}$ where the ordering is reversed. Recalling
 that the cut is given by multiplication of two tree-level
 amplitudes, where one has $5$ legs and the other has $6$,
 and using the identity \eqn\REFIDE{ A_{n}^{\rm tree}(1,2,..., n)  =
(-)^n A_{n}^{\rm tree} (n,...,2,1) } we immediately get the
following results. If the cut $C_{456}$ is given by some function
$f(1,2,3,4,5,6,7)$, then the cut $C_{567}$ is given by
$-f(3,2,1,7,6,5,4)$. Since the cut structure determines the
amplitude completely, the same reflection property holds for the amplitude as well. Now,
remember that the amplitude can be expanded 
 into box functions as $\sum_{j} a_j F_{ j}$,
where $F_j$ represents all the  box functions. If
the action of $\sigma$ on indices transforms  $F_{
k}\rightarrow F_{ t}$, we find immediately that 
 $a_t|=- a_k|_{\sigma}$, where $|_{\sigma}$ means 
to act $\sigma$ on the gluon labels in the function $a_k$. 
For our example, we have \eqn\RELA{\eqalign{ b_1 & = -
b_7|_{\sigma},~~b_2  = - b_6|_{\sigma},~~b_3  = - b_5|_{\sigma}, ~~b_4  = -
b_4|_{\sigma},\cr c_1 & = - c_2|_{\sigma},~~c_3  = - c_7|_{\sigma},~~c_4 =
- c_6|_{\sigma}, ~~c_5 = - c_5|_{\sigma},\cr g_1 & = - g_5|_{\sigma},~~ g_2
=  -g_4|_{\sigma},~~ g_3 = - g_3|_{\sigma},~~ g_6 = - g_7|_{\sigma},~~~~\cr
d_{2,1} & = -d_{3,6}|_{\sigma},~~d_{2,2} = -d_{3,5}|_{\sigma},~~d_{2,3}
= -d_{3,4}|_{\sigma},~~d_{2,4} = -d_{3,3}|_{\sigma},\cr d_{2,5} &  =
-d_{3,2}|_{\sigma},~~d_{2,6} = -d_{3,1}|_{\sigma},~~d_{2,7} =
-d_{3,7}|_{\sigma}. }}
Applying this transformation to the coefficients we have already
computed yields expressions for the following previously
undetermined coefficients:
$$ b_1,~b_5,~c_3,~c_6,~d_{2,3},~d_{2,7}~d_{3,2},~d_{3,6},~g_3. $$
The explicit expressions are listed in section 4.

% % % % % % % % % % % %
\subsec{Completion And Consistency Checks}
% % % % % % % % % % % % 

At this point we have succeeded in computing thirty-one of the
coefficients. In principle we could compute the remaining four
coefficients by applying the same general method of section 2 to the
remaining two cuts, i.e., $C_{345}$ and $C_{671}$.

The four coefficients we are missing are $b_2$, $b_6$, $d_{2,4}$
and $d_{3,3}$.

{}From the condition that both $C_{345}$ and $C_{671}$ be finite,
we obtain two equations:
\eqn\kilo{ \eqalign{ -b_6+d_{2,4}-{1\over 2}d_{3,3} =&~
-c_{3}-c_4+c_7+d_{3,7}-{1\over 2}\left( -d_{2,1}+g_4+g_6\right),
\cr -b_2-{1\over 2}d_{2,4}+d_{3,3} =&~ c_3-c_6-c_7-d_{2,7}-{1\over
2}\left( -d_{3,6}+g_2+g_7 \right). }}
Therefore we are left with the problem of determining two
coefficients, say $b_2$ and $b_6$.

Before we derive the remaining coefficients, let us make some
observations about the known infrared singular behavior of
one-loop amplitudes \refs{\giel, \kuni}. We have already found
that in the final form of the amplitude all singular terms of the
form
\eqn\three{ -{1\over \epsilon^2}\left( -t_i^{[3]}
\right)^{-\epsilon}}
cancel for all $i =1 ,\ldots , 7$. This is the statement that
cuts in three-particle channels are finite. However, up to now we have not
considered cuts in two-particle channels. It turns out that the singular
behavior in these cuts is universal and produces a term in the
amplitude of the form
\eqn\lead{\eqalign{ & A^{\rm
1-loop}_{7:1}(1^-,2^-,3^-,4^+,5^+,6^+,7^+)|_{\rm IR} = \cr &
\left[ -{1\over \epsilon^2}\sum_{i=1}^7 \left(
-t_i^{[2]}\right)^{-\epsilon} \right]A^{\rm
tree}_7(1^-,2^-,3^-,4^+,5^+,6^+,7^+). }}
Note that this translates into seven equations our coefficients
have to satisfy.  

Taking the terms of \lead\ involving the $i=5$ singularity,
 we find that our coefficients have to
satisfy the following equation (see Appendix A for details of the derivation):
\eqn\woop{ b_1+b_7-c_5+{1\over 2}\left( -d_{2,5}+d_{2,7}-d_{3,2}+
d_{3,7} - g_1 -g_5 \right) = A^{\rm
tree}_{7}(1^-,2^-,3^-,4^+,5^+,6^+,7^+).}
This equation only involves known coefficients and is therefore
a consistency check.

The tree-level seven-gluon amplitude is given by \seventree
\eqn\ATREE{\eqalign{ & ~~A^{\rm tree}(1^-,2^-,3^-,4^+,5^+,6^+,7^+) \cr
& =\left[{\vev{2~3} \gb{1| 6+7|5}  \gb{1|2+3|4}^2 \over
[2~3]\vev{5~6}\vev{6~7}\vev{7~1} t_3^{[2]} t_2^{[3]}
t_6^{[3]}} -{\vev{2~1} \gb{3|5+4|6} \gb{3|2+1|7}^2 \over [2~1]
\vev{6~5} \vev{5~4} \vev{4~3} t_7^{[2]} t_7^{[3]} t_3^{[3]}
}\right] \cr & +\left[ {[4~5] \vev{1~2}\gb{3|1+2|7}
(\vev{5~6}\gb{3|1+2|6} +\vev{5~7} \gb{3|1+2|7} )\over
[1~2]\vev{4~5}\vev{5~6}\vev{6~7} t_3^{[2]}t_7^{[2]} t_3^{[3]}}
\right.\cr & \left. - { [7~6]\vev{3~2} \gb{1|3+2|4}(\vev{6~5}
\gb{1|3+2|5}+\vev{6~4} \gb{1|3+2|4}) \over [3~2] \vev{7~6}
\vev{6~5}\vev{5~4} t_7^{[2]}t_3^{[2]} t_6^{[3]}}\right] \cr &
+\left[{ \vev{1~2}\vev{2~3} [4~5][6~7] ( (\vev{3~4}[6~4]\vev{1~6}-
\vev{1~7}[5~7]\vev{3~5})+(
\vev{3~4}[7~4]\vev{1~7})+(\vev{1~6}[6~5]\vev{3~5})) \over \vev{4~5}
\vev{6~7} t_3^{[2]} t_7^{[2]} t_3^{[3]} t_6^{[3]}}
\right]\cr & + \left[{ \gb{1|2+3|4} \gb{3|2+1|7} t_1^{[3]}
\over[1~2][2~3]
 t_7^{[2]} t_3^{[2]} \vev{4~5}\vev{5~6}\vev{6~7}}\right] .
}}

With the help of a symbolic manipulation program, we have analytically 
verified the relation \woop.  From the
form of the seven-gluon tree amplitude \ATREE\ it is clear that this is an impressive check
of our coefficients.

Now that we have checked our previous calculations, we can use two
of the equations in \lead\ that involve the unknown coefficients,
i.e., $b_2$ and $b_6$, in order to find them. Take for example the
equations derived from looking at the $i=4$ and $i=7$ terms in
\lead,
\eqn\unki{\eqalign{ b_6+b_7-c_4-{1\over 2}\left(
d_{2,4}-d_{2,6}+d_{3,1}-d_{3,6}+g_4+g_7\right) = & ~ A^{\rm
tree}_{7}(1^-,2^-,3^-,4^+,5^+,6^+,7^+), \cr b_2+b_3-c_7+{1\over
2}\left( d_{2,2}-d_{2,7}+d_{3,2}-d_{3,4}-g_3-g_7 \right) = & ~
A^{\rm tree}_{7}(1^-,2^-,3^-,4^+,5^+,6^+,7^+). }}
These two equations give $b_6$ and $b_2$ in terms of known
coefficients respectively. 
They are expressed as
\eqn\defb{\eqalign{b_6 =& ~ A^{\rm
tree}_{7}(1^-,2^-,3^-,4^+,5^+,6^+,7^+)-b_7+c_4+ {1\over 2}\left(
d_{2,4}-d_{2,6}+d_{3,1}-d_{3,6}+g_4+g_7\right), \cr  b_2 = &~
A^{\rm tree}_{7}(1^-,2^-,3^-,4^+,5^+,6^+,7^+) -b_3+c_7- {1\over
2}\left( d_{2,2}-d_{2,7}+d_{3,2}-d_{3,4}-g_3-g_7 \right).}}

Finally, using  these expressions for $b_2$ and
$b_6$ in the two equations in \kilo, we solve for $d_{2,4}$ and
$d_{3,3}$ to find
\eqn\lastwo{\eqalign{ d_{2,4}=&~ 2 A^{\rm tree}_7 - 2b_4-2
b_5+d_{2,2}-d_{3,4}+d_{3,6}+g_5, \cr  d_{3,3} =&~ 2 A^{\rm tree}_7
-2b_4-2b_3+d_{3,5}-d_{2,3}+d_{2,1}+g_1. } }

This completes the list of all thirty-five coefficients in the
one-loop seven-gluon amplitude.

Now we use the remaining equations derived from the infrared
structure \lead\ as further consistency checks of our coefficients. We
successfully checked that the equations for $i=1,2,3,6$ are
satisfied.

In the next section we summarize our results.

%%%%%%%%%%%%%%%%%%%%%%%%%%
\newsec{The Full Amplitude $A_{7;1}(1^-, 2^-, 3^-, 4^+, 5^+, 6^+, 7^+)$}
%%%%%%%%%%%%%%%%%%%%%%%%%%

Here  we summarize all results for $A_{7;1}(1^-, 2^-, 3^-,
4^+, 5^+, 6^+, 7^+)$ that were scattered through the previous
sections into one complete form, so that a reader interested only
in results can skip all derivations. The amplitude is
\eqn\TOTAL{\eqalign{ A_{7;1}(1^-, 2^-, 3^-, 4^+, 5^+, 6^+, 7^+)
& = \sum_{j=1}^7 b_j F^{1m}_{7:j} +\sum_{j=1}^7 c_j
F^{2m~e}_{7:2;j} +\sum_{j=1}^7 d_{2,j} F^{2m~h}_{7:2;j} \cr & +
\sum_{j=1}^7 d_{3,j}
 F^{2m~h}_{7:3;j}+ \sum_{j=1}^7 g_j F^{3m}_{7:2:2;j}
}} A few remarks must be made before we list the thirty-five
coefficients. Twenty-five of them have explicit forms.  Four of
them ($b_1,b_3,b_5,b_7$) are expressed in terms of the twenty-five
explicit ones.  We stress that we calculated them independently
but are abbreviating them for convenience only.  The last six
coefficients were derived in terms of the others in the following
order: $d_{2,7},~d_{3,7},~b_2,~b_6,~d_{2,4},~d_{3,3}$.

First we recall our conventions and make a couple of convenient
definitions:
 \eqn\tosimy{\eqalign{
2 p_i \cdot p_j &= \vev{i~j} [i~j], \cr t_i^{[r]} &=
(p_i+p_{i+1}+\cdots+p_{i+r-1})^2, \cr \gb{i|j_r+j_{r+1}+\cdots
+j_s|k} &\equiv
\vev{i~j_r}[j_r~k]+\vev{i~j_{r+1}}[j_{r+1}~k]+\cdots+\vev{i~j_s}[i~j_s],\cr
  S_1 & \equiv {\gb{ 3| 1+2|7}^3
\over t_7^{[3]}  [7~1][1~2]\vev{3~4}\vev{4~5} \vev{5~6} \gb{ 6| 7+1|
2}}, \cr S_2 & \equiv -{ \gb{1|3+2| 4}^3\over t_{2}^{[3]}
[4~3][3~2] \vev{1~7}\vev{7~6} \vev{6~5} \gb{ 5| 4+3| 2}}. }}

Here is the list of the thirty-five coefficients.
\eqn\RELAA{\eqalign{ b_1 &= c_6+ g_6, \cr b_2 & = A^{\rm tree}-
b_3+c_7 -{1\over 2} d_{2,2}+{1\over 2} d_{2,7} -{1\over 2}
d_{3,2}+{1\over 2} d_{3,4}+{1\over 2} g_3+{1\over 2} g_7\cr b_3 &
= g_1+ d_{2,1}, \cr b_4 &= { (t_1^{[3]})^3 \over
[1~2][2~3]\vev{4~5}\vev{5~6}\vev{6~7}\gb{4|2+3|1}\gb{7|1+2|3}},
\cr b_5 & = g_5+ d_{3,6}, \cr b_6 & = A^{\rm tree} -b_5+c_3 -{1\over
2}d_{3,5}+{1\over 2}d_{3,7}-{1\over 2} d_{2,5}+{1\over 2}d_{2,3}+
{1\over 2} g_3 +  {1\over 2} g_6, \cr b_7 &= c_4 + g_7 .
 }}

\eqn\RELAE{\eqalign{ 
c_1 &=0, \cr c_2 &=0, \cr c_3 & = {
\vev{2~3}^3 [6~7]^3 \over t_{6}^{[3]} \vev{3~4} \vev{4~5} [7~1] \gb{ 2|
7+1|6} \gb{5|6+7| 1}} \cr c_4 & = S_1 \cr c_5 & = {
(t_{1}^{[3]})^2\over [1~2][3~2] \vev{4~5}\vev{7~6}\vev{5~6} \gb{4|
2+3|1} \gb{7|2+1|3}} \cr c_6 & = S_2 \cr c_7 & = - { \vev{2~1}^3
[5~4]^3 \over t_{3}^{[3]} \vev{1~7} \vev{7~6} [4~3] \gb{2| 4+3| 5}
\gb{6| 5+4|3}} }}

\eqn\RELAC{\eqalign{ 
d_{2,1} & = S_1 - { \gb{ 3| 4+5|6}^3
\vev{1~2}^3 \over \vev{7~1}\vev{3~4}\vev{4~5} t_{3}^{[3]} t_{7}^{[3]}
\gb{ 2|7+1|6} ( \vev{6~5} \gb{7|1+2| 6}-\vev{7~5} t_{7}^{[3]})} \cr
d_{2,2} &= { (t_1^{[3]})^3 \over
[1~2][2~3]\vev{4~5}\vev{5~6}\vev{6~7}\gb{4|2+3|1}\gb{7|1+2|3}},
\cr 
d_{2,3} &= S_2- { \vev{3~2}^3 (t_{5}^{[3]})^3 \over \vev{3~4} \vev{5~6}
\vev{6~7} t_{2}^{[3]} \gb{ 4| 2+3 |1} \gb{5|6+7|1} (  t_{5}^{[4]}
\vev{2~7} -\vev{1~7} \gb{ 2|3+4|1})} \cr 
d_{2,4} & = 2 A^{\rm tree} -2
b_4-2b_5 + d_{2,2} -d_{3,4} +d_{3,6} + g_5 \cr 
d_{2,5} & = {
\vev{1~2}^3 (t_{4}^{[3]})^3 \over \vev{4~5}\vev{5~6}
\vev{7~1} t_{7}^{[3]} \gb{7|1+2|3} \gb{6|4+5|3}
(\vev{4~2}t_{4}^{[3]} +\vev{2~3} \gb{4|5+6|3})} \cr 
d_{2,6} &= 0 \cr
d_{2,7} &= -2 b_5+ 2c_3-2c_6 +2 d_{2,3} -d_{3,2}+ 2d_{3,6}+ g_3+
g_5. }}

\eqn\RELAD{\eqalign{ 
d_{3,1} &= 0, \cr 
d_{3,2} & = - {
\vev{3~2}^3 (t_{5}^{[3]})^3 \over \vev{7~6}\vev{6~5} \vev{4~3}
t_{2}^{[3]}  \gb{4|3+2|1} \gb{5|7+6|1}(\vev{7~2}t_{5}^{[3]}
+\vev{2~1}\gb{7|6+5|1})} \cr d_{3,3} & = 2 A^{\rm tree} -2 b_4 -2 b_3+
d_{3,5} -d_{2,3} + d_{2,1}+ g_1\cr 
d_{3,4} & = S_1-
{\vev{1~2}^3 (t_{4}^{[3]})^3 \over \vev{4~5} \vev{5~6} \vev{7~1} t_{7}^{[3]} \gb{
7| 1+2 |3} \gb{6| 4+5|3} (  t_{3}^{[4]} \vev{2~4} -\vev{3~4} \gb{
2|7+1| 3})} \cr d_{3,5} &=  { (t_1^{[3]})^3 \over
[1~2][2~3]\vev{4~5}\vev{5~6}\vev{6~7}\gb{4|2+3|1}\gb{7|1+2|3}},
\cr d_{3,6} & = S_2 + { \gb{1| 7+6|5}^3 \vev{3~2}^3\over
\vev{4~3}\vev{1~7}\vev{7~6} t_{6}^{[3]} t_{2}^{[3]} \gb{2|4+3|5}
(\vev{5~6}\gb{4|3+2| 5}-\vev{4~6} t_{2}^{[3]})} \cr d_{3,7} &= -2
b_3+ 2c_7-2c_4+2d_{3,4}-d_{2,5}+2d_{2,1}+ g_3+ g_1 }}

\eqn\RELAG{\eqalign{ 
g_1 & = { \vev{1~2}^3 \gb{7 | 5+6|4}^3 \over
\vev{5~6} \vev{6~7} \vev{7~1} [3~4] \gb{7| 1+2| 3} ( \vev{7~2}
t_5^{[3]} +\vev{2~1} \gb{7|5+6|1}) (\vev{6~5}  \gb{7| 1+2|
6}-\vev{7~5} t_7^{[3]})}\cr 
g_2 &=0 \cr 
g_3 & = - { \vev{1~2}^3
\vev{2~3}^3 [5~6]^3 \over \vev{7~1} \vev{3~4} \gb{2 | 3+4| 5} \gb{ 2|
7+1|6} ( \vev{7~1} \gb{ 2| 3+4 |1} - t_{2}^{[3]} \vev{7~2}) (
t_{3}^{[4]}\vev{2~4} -\vev{3~4} \gb{ 2|7+1| 3})} \cr 
g_4 &= 0 \cr
g_5 & = -{\vev{3~2}^3 \gb{4|6+5|7}^3\over \vev{6~5} \vev{5~4}
\vev{4~3} [1~7] \gb{4| 3+2|1} (\vev{4~2}t_4^{[3]} +\vev{2~3}
\gb{4|6+5|3}) (\vev{5~6} \gb{4| 3+2| 5}-\vev{4~6} t_2^{[3]})} \cr
g_6 & = { (\vev{5~3}  t_5^{[3]}-\vev{4~3} \gb{5|6+7|4})^3 \over [1~2]
\vev{3~4}\vev{4~5}\vev{5~6}\vev{6~7}\gb{5|6+7|1} \gb{5|3+4|2}
(\vev{7~5}  t_3^{[3]}-\vev{7~6}\gb{5|3+4|6})}\cr 
g_7 & = -
{(\vev{6~1}  t_4^{[3]}-\vev{7~1} \gb{6|5+4|7})^3 \over [3~2]
\vev{1~7}\vev{7~6}\vev{6~5}\vev{5~4}\gb{6|5+4|3} \gb{6|1+7|2}
(\vev{4~6}  t_6^{[3]}-\vev{4~5}\gb{6|1+7|5})} }}

We repeat here the tree-level amplitude \seventree\ for the reader's convenience.
\eqn\ATREE{\eqalign{ & ~~A^{\rm tree}(1^-,2^-,3^-,4^+,5^+,6^+,7^+) \cr
& =\left[{\vev{2~3} \gb{1| 6+7|5}  \gb{1|2+3|4}^2 \over
[2~3]\vev{5~6}\vev{6~7}\vev{7~1} t_3^{[2]} t_2^{[3]}
t_6^{[3]}} -{\vev{2~1} \gb{3|5+4|6} \gb{3|2+1|7}^2 \over [2~1]
\vev{6~5} \vev{5~4} \vev{4~3} t_7^{[2]} t_7^{[3]} t_3^{[3]}
}\right] \cr & +\left[ {[4~5] \vev{1~2}\gb{3|1+2|7}
(\vev{5~6}\gb{3|1+2|6} +\vev{5~7} \gb{3|1+2|7} )\over
[1~2]\vev{4~5}\vev{5~6}\vev{6~7} t_3^{[2]}t_7^{[2]} t_3^{[3]}}
\right.\cr & \left. - { [7~6]\vev{3~2} \gb{1|3+2|4}(\vev{6~5}
\gb{1|3+2|5}+\vev{6~4} \gb{1|3+2|4}) \over [3~2] \vev{7~6}
\vev{6~5}\vev{5~4} t_7^{[2]}t_3^{[2]} t_6^{[3]}}\right] \cr &
+\left[{ \vev{1~2}\vev{2~3} [4~5][6~7] ( (\vev{3~4}[6~4]\vev{1~6}-
\vev{1~7}[5~7]\vev{3~5})+(
\vev{3~4}[7~4]\vev{1~7})+(\vev{1~6}[6~5]\vev{3~5})) \over \vev{4~5}
\vev{6~7} t_3^{[2]} t_7^{[2]} t_3^{[3]} t_6^{[3]}}
\right]\cr & + \left[{ \gb{1|2+3|4} \gb{3|2+1|7} t_1^{[3]}
\over[1~2][2~3]
 t_7^{[2]} t_3^{[2]} \vev{4~5}\vev{5~6}\vev{6~7}}\right] .
}}

We have written the tree amplitude so that every bracketed expression changes sign under the index
shift $1\leftrightarrow 3, 4\leftrightarrow 7, 5\leftrightarrow
6$. This is the reflection symmetry made manifest.

%%%%%%%%%%%%%%%%%%%%%%%%%%%%%%%%%%
\bigskip
\bigskip

\centerline{\bf Acknowledgements}

We thank O. Lunin and P. Svr\v cek for helpful discussions and D. Kosower for a question prompting us to add Appendix B.
R.B. and B.F. were supported by NSF grant PHY-0070928.
F.C. was supported in part by the Martin A. and Helen Chooljian Membership at the Institute for Advanced Study and  by DOE grant DE-FG02-90ER40542.

%%%%%%%%%%%%%%%%%%%%%%%%%%%%%%%%%%%
\appendix{A}{Box Functions and Divergence Analysis}

The scalar box functions used in this paper are the following:
\eqn\expi{\eqalign{ F^{1m}_{n:i} = & ~ -{1\over \epsilon^2}\left[
(-t_{i-3}^{[2]})^{-\epsilon} +(-t_{i-2}^{[2]})^{-\epsilon}
-(-t_{i-3}^{[3]})^{-\epsilon} \right] \cr & + {\rm Li}_2 \left( 1-
{t_{i-3}^{[3]}\over t_{i-3}^{[2]}} \right) + {\rm Li}_2 \left( 1-
{t_{i-3}^{[3]}\over t_{i-2}^{[2]}} \right) + {1\over 2}\ln^2\left(
{t_{i-3}^{[2]}\over t_{i-2}^{[2]}} \right) + {\pi^2\over 6},}}
\eqn\exop{\eqalign{ F^{2m~e}_{n:r;i} = & ~ -{1\over
\epsilon^2}\left[ (-t_{i-1}^{[r+1]})^{-\epsilon}
+(-t_{i}^{[r+1]})^{-\epsilon} -(-t_{i}^{[r]})^{-\epsilon} -
(-t_{i-1}^{[r+2]})^{-\epsilon} \right] \cr & + {\rm Li}_2 \left(
1- {t_{i}^{[r]}\over t_{i-1}^{[r+1]}} \right) + {\rm Li}_2 \left(
1- {t_{i}^{[r]}\over t_{i}^{[r+1]}} \right) +{\rm Li}_2 \left( 1-
{t_{i-1}^{[r+2]}\over t_{i-1}^{[r+1]}} \right) \cr & + {\rm Li}_2
\left( 1- {t_{i-1}^{[r+2]}\over t_{i}^{[r+1]}} \right) - {\rm
Li}_2 \left( 1- {t_{i}^{[r]}t_{i-1}^{[r+2]}\over
t_{i-1}^{[r+1]}t_{i}^{[r+1]}} \right) + {1\over 2}\ln^2\left(
{t_{i-1}^{[r+1]}\over t_{i}^{[r+1]}} \right),}}
\eqn\uxop{\eqalign{ F^{2m~h}_{n:r;i} = & ~ -{1\over
\epsilon^2}\left[ (-t_{i-2}^{[2]})^{-\epsilon}
+(-t_{i-1}^{[r+1]})^{-\epsilon} -(-t_{i}^{[r]})^{-\epsilon} -
(-t_{i-2}^{[r+2]})^{-\epsilon} \right] \cr & -{1\over
2\epsilon^2}{(-t_{i}^{[r]})^{-\epsilon}(-t_{i-2}^{[r+2]})^{-\epsilon}
\over (-t_{i-2}^{[2]})^{-\epsilon}}  + {1\over 2}\ln^2 \left(
{t_{i-2}^{[2]}\over t_{i-1}^{[r+1]}} \right) \cr & +  {\rm Li}_2
\left( 1- {t_{i}^{[r]}\over t_{i-1}^{[r+1]}} \right) + {\rm Li}_2
\left( 1- {t_{i-2}^{[r+2]}\over t_{i-1}^{[r+1]}} \right), }}
\eqn\uliz{\eqalign{ F^{3m}_{n:r:r';i} = & ~ -{1\over
\epsilon^2}\left[ (-t_{i-1}^{[r+1]})^{-\epsilon}
+(-t_{i}^{[r+r']})^{-\epsilon} -(-t_{i}^{[r]})^{-\epsilon} -
(-t_{i+r}^{[r']})^{-\epsilon} - (-t_{i-1}^{[r+r'+1]})^{-\epsilon}
\right] \cr & -{1\over
2\epsilon^2}{(-t_{i}^{[r]})^{-\epsilon}(-t_{i+r}^{[r']})^{-\epsilon}
\over (-t_{i}^{[r+r']})^{-\epsilon}} -{1\over
2\epsilon^2}{(-t_{i+r}^{[r']})^{-\epsilon}(-t_{i-1}^{[r+r'+1]})^{-\epsilon}
\over (-t_{i-1}^{[r+1]})^{-\epsilon}} + {1\over 2}\ln^2 \left(
{t_{i-1}^{[r+1]}\over t_{i}^{[r+r']}} \right) \cr & +  {\rm Li}_2
\left( 1- {t_{i}^{[r]}\over t_{i-1}^{[r+1]}} \right) + {\rm Li}_2
\left( 1- {t_{i-1}^{[r+r'+1]}\over t_{i}^{[r+r']}} \right) - {\rm
Li}_2 \left( 1- {t_i^{[r]}t_{i-1}^{[r+r'+1]}\over
t_{i-1}^{[r+1]}t_{i}^{[r+r']}} \right), }}
The dilogarithm function is defined by ${\rm Li}_2(x) = -\int_0^x \ln
(1-z)dz/z$.
Now we specialize to seven gluons and 
discuss the infrared singular structure of the one-loop amplitude.  
Recall that the seven-gluon amplitude is written as a sum of scalar box functions as in \bcddg.  The box functions contain divergences when $\epsilon \rightarrow 0$ of the form 
\eqn\mmpovu{{1 \over \epsilon^2} (-t_{i}^{[2]})^{-\epsilon},~~~~~~~~~~~~
{1 \over \epsilon^2} (-t_{i}^{[3]})^{-\epsilon},}
remembering that $t_i^{[r]}=t_{i+r}^{[7-r]}$ for seven gluons, by momentum conservation.

Now it is clear that the divergent structure of the seven-gluon amplitude takes the form
\eqn\POLEgen{
A_{7:1}|_{\rm IR}= -{1\over \epsilon^2} 
\sum_{i=1}^7 \left( \alpha_i 
(-t_i^{[2]})^{-\epsilon}+\beta_i 
(-t_i^{[3]})^{-\epsilon}\right),
}
where $\alpha_i$ and $\beta_i$ are linear combinations of the coefficients in \bcddg.  The $\alpha_i$ and $\beta_i$ appear in the body of the paper.  Here we describe how to compute them from the box functions, taking $\alpha_5$ as an example.

The infrared behavior of the box functions contributing to 
 $\alpha_5$ are as follows.
\eqn\alffrombox{\eqalign{
F_{7:1}^{1m}|_{\rm IR} &= -{1\over \epsilon^2}\left[
(-t_{5}^{[2]})^{-\epsilon} +(-t_{6}^{[2]})^{-\epsilon}
-(-t_{5}^{[3]})^{-\epsilon} \right] \cr
F_{7:7}^{1m}|_{\rm IR} &= -{1\over \epsilon^2}\left[
(-t_{4}^{[2]})^{-\epsilon} +(-t_{5}^{[2]})^{-\epsilon}
-(-t_{4}^{[3]})^{-\epsilon} \right] \cr
F_{7:2;5}^{2m~e}|_{\rm IR} &= -{1\over
\epsilon^2}\left[ (-t_{4}^{[3]})^{-\epsilon}
+(-t_{5}^{[3]})^{-\epsilon} -(-t_{5}^{[2]})^{-\epsilon} -
(-t_{1}^{[3]})^{-\epsilon} \right] \cr
F_{7:2;5}^{2m~h}|_{\rm IR} &= -{1\over
\epsilon^2}\left[ \half(-t_{3}^{[2]})^{-\epsilon}
+(-t_{4}^{[3]})^{-\epsilon} -\half(-t_{5}^{[2]})^{-\epsilon} -
\half(-t_{7}^{[3]})^{-\epsilon} \right] \cr
F_{7:2;7}^{2m~h}|_{\rm IR} &= -{1\over
\epsilon^2}\left[ \half(-t_{5}^{[2]})^{-\epsilon}
+(-t_{6}^{[3]})^{-\epsilon} -\half(-t_{7}^{[2]})^{-\epsilon} -
\half(-t_{2}^{[3]})^{-\epsilon} \right] \cr
F_{7:2:2;1}^{3m}|_{\rm IR} &=  -{1\over
\epsilon^2}\left[ \half(-t_{7}^{[3]})^{-\epsilon}
+\half(-t_{5}^{[3]})^{-\epsilon} -\half(-t_{1}^{[2]})^{-\epsilon} 
 - \half(-t_{5}^{[2]})^{-\epsilon}
\right] \cr
F_{7:2:2;5}^{3m}|_{\rm IR} &=  -{1\over
\epsilon^2}\left[ \half(-t_{4}^{[3]})^{-\epsilon}
+\half(-t_{2}^{[3]})^{-\epsilon} -\half(-t_{5}^{[2]})^{-\epsilon} 
 - \half(-t_{2}^{[2]})^{-\epsilon}
\right] \cr
}}
 Collecting all the terms with $t_5^{[2]},$
 we find that
\eqn\ALPHAFIVE{
\alpha_5= b_1+b_7 -c_5 -{1\over 2} d_{2,5}+{1\over 2} d_{2,7}
-{1\over 2} d_{3,2}+{1\over 2} d_{3,7} -{1\over 2} g_1-{1\over 2} g_5.
}
Similar calculations give expressions for the $\beta_i.$

%%%%%%%%%%%%%%%%%%%%%%%%%%%%%%%%%%%%%%
\appendix{B}{Constructing a General Next-to-MHV Amplitude}

Here we flesh out the claim that our method can compute all next-to-MHV amplitudes (i.e.~those with exactly three negative helicities, in arbitrary positions).  In section 2 we already argued that all next-to-MHV amplitudes have the property, required for our method, that one of the tree-level amplitude factors in the cut integral \cutIn\ is MHV.  But to calculate the amplitude, we must be sure that we can determine each coefficient in \gene\ from one of these cuts.  

To see that this is correct, consider the scalar box function associated to each coefficient.  To be able to determine the coefficient by our method, the box function must have the property that it appears in some cut $C_{i,i+1,\ldots,j}$, where the amplitude $A^{\rm tree}((-\ell_1),i,i+1,\ldots ,j-1,j,(-\ell_2))$ is MHV, but is not annihilated by all operators $[F_{klm},\eta]$ where $i\leq k,l,m \leq j$.  (The operator $[F_{klm},\eta]$ annihilates box functions where gluons $k,l,m$ are attached to the same corner of the box.)

One-mass scalar box functions appear in only one cut (disregarding cuts in two-particle channels).  See Figure 2.  The cut has three gluons, say $k,l,m$, on one side.  Since the tree-level amplitude on that side has five particles, it is MHV (unless it vanishes).  The box function appears in the cut $C_{klm}$ and is not annihilated by the operator $[F_{klm},\eta]$, so the coefficient can be calculated by this operator acting on this cut.

\ifig\allntmhv{Cuts of a general next-to-MHV amplitude.  The first row shows cuts, for each type of scalar box function, that are guaranteed to have at least three gluons on each side.  The second row illustrates, for two-mass and three-mass box functions, how to identify one side or the other as an MHV tree amplitude, so that a suitable operator can be chosen to calculate the coefficient.}
{\epsfxsize=1.0\hsize\epsfbox{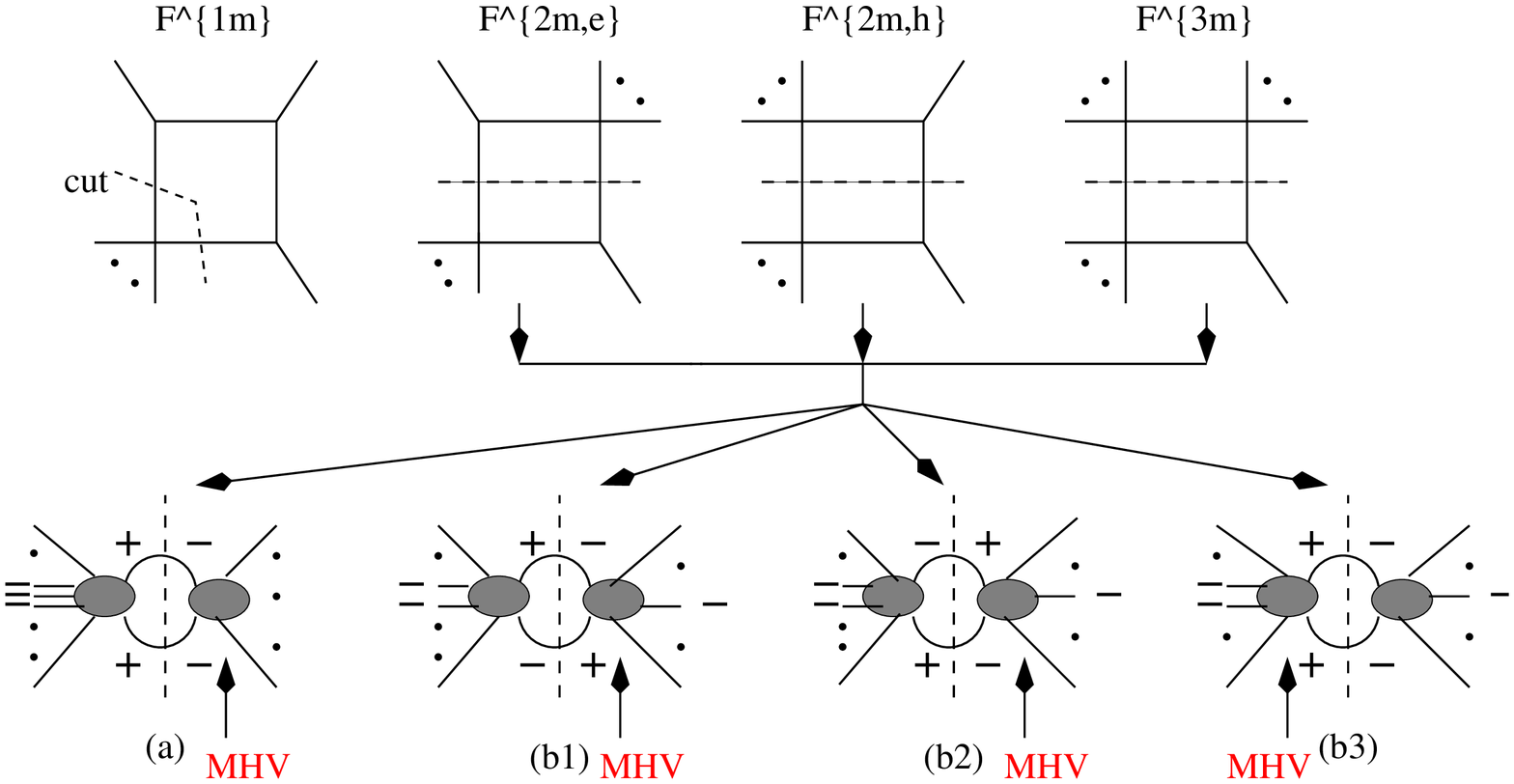}}

The remaining scalar box functions can be analyzed as a group.  For this general analysis, we should consider the cuts indicated in Figure 2, to be sure that there are at least three gluons on each side.  There are two cases.  Case (a): If all three of the negative-helicity gluons appear on the same side of the cut, then the opposite side must be MHV (or vanish).  We can choose $k,l,m$ from that side such that they are not all on the same corner of the box.  Case (b):  If there are two negative-helicity gluons on one side of the cut, and one on the other, then the sides will alternately be MHV, depending on the helicity assignments of the cut propagators.  In any case it is possible to choose three gluons $k,l,m$ from the MHV side that are not all on the same corner of the box.  The operator $[F_{klm},\eta]$ then can be used to analyze the cut in question without annihilating the scalar box function.  The three
cases (b1), (b2), (b3) would suggest using two separate operators,
depending on which side of the cut is MHV.  In fact, a single
one of them will suffice to determine the coefficient of a particular box
function.

%%%%%%%%%%%%%%%%%%%%%%%%%%%%%%%%%%%%%%%%
\listrefs

\end